\documentclass[lettersize,journal]{IEEEtran}
\usepackage{amsmath,amsfonts}
\usepackage{algorithmic}
\usepackage{array}
\usepackage[caption=false,font=normalsize,labelfont=sf,textfont=sf]{subfig}
\usepackage{textcomp}
\usepackage{stfloats}
\usepackage{url}
\usepackage{verbatim}
\usepackage{graphicx}

\def\BibTeX{{\rm B\kern-.05em{\sc i\kern-.025em b}\kern-.08em
    T\kern-.1667em\lower.7ex\hbox{E}\kern-.125emX}}
\usepackage{balance}

\mathchardef\mhyphen="2D
\graphicspath{ {figures/} }
\usepackage{cleveref}
\usepackage{xcolor}
\usepackage{multirow}
\usepackage{xcolor}

\newcommand{\modelp}{\emph{MP}}
\newcommand{\classp}{\emph{CP}}
\newcommand{\variantp}{\emph{VP}}

\definecolor{darkred}{HTML}{8B0000}
\newcommand{\revise}[2]{%
	\ifcase#1%
		\textcolor{black}{#2}% No revision
	\or
		\textcolor{black}{#2}% Revision 1
	\or
		\textcolor{black}{#2}% Revision 2
	\fi
}

\usepackage{tikz}
\newcommand\copyrighttext{%
  \footnotesize \textcopyright~\the\year~IEEE. Personal use of this material is permitted.
  Permission from IEEE must be obtained for all other uses, in any current or future
  media, including reprinting/republishing this material for advertising or promotional
  purposes, creating new collective works, for resale or redistribution to servers or
  lists, or reuse of any copyrighted component of this work in other works.}
\newcommand\copyrightnotice{%
\begin{tikzpicture}[remember picture,overlay]
\node[anchor=north,yshift=-15pt,xshift=-10pt] at (current page.north) {\fbox{\parbox{\dimexpr\textwidth-\fboxsep-\fboxrule\relax}{\copyrighttext}}};
\end{tikzpicture}
\vspace{-0.3cm}
}
\begin{document}
\title{Variant Parallelism: Lightweight Deep Convolutional Models for Distributed Inference on IoT Devices}
\author{Navidreza Asadi, Maziar Goudarzi
\thanks{
Navidreza Asadi is now with Computer Engineering Department, Technical University of Munich, Germany (navidreza.asadi@tum.de). He was with the Sharif University of Technology while working on this project. Maziar Goudarzi is with the Computer Engineering Department, Sharif University of Technology, Tehran, Iran. (goudarzi@sharif.edu);}}
% \thanks{Manuscript created October, 2020; This work was developed by the IEEE Publication Technology Department. This work is distributed under the \LaTeX \ Project Public License (LPPL) ( http://www.latex-project.org/ ) version 1.3. A copy of the LPPL, version 1.3, is included in the base \LaTeX \ documentation of all distributions of \LaTeX \ released 2003/12/01 or later. The opinions expressed here are entirely that of the author. No warranty is expressed or implied. User assumes all risk.}}

% \markboth{Journal of \LaTeX\ Class Files,~Vol.~18, No.~9, September~2020}%
% \markboth{Submitted to IEEE Internet of Things Journal}%
\markboth{}%
% \markboth{}
{Variant Parallelism}

\maketitle
\copyrightnotice

\begin{abstract}
  Two major techniques are commonly used to meet real-time inference limitations when distributing models across resource-constrained IoT devices: (1) model parallelism (\modelp{}) and (2) class parallelism (\classp{}). In \modelp{}, transmitting bulky intermediate data (orders of magnitude larger than input) between devices imposes huge communication overhead. Although \classp{}  solves this problem, it has limitations on the number of sub-models. In addition, both solutions are fault intolerant, an issue when deployed on edge devices. We propose variant parallelism (\variantp{}), an ensemble-based deep learning distribution method where different variants of a main model are generated and can be deployed on separate machines. We design a family of lighter models around the original model, and train them simultaneously to improve accuracy over single models. Our experimental results on six common mid-sized object recognition datasets demonstrate that our models can have $\boldsymbol{5.8\mhyphen 7.1\times}$ fewer parameters, $\boldsymbol{4.3\mhyphen 31\times}$ fewer multiply-accumulations (MACs), and $\boldsymbol{2.5\mhyphen 13.2\times}$ less response time on atomic inputs compared to MobileNetV2 while achieving comparable or higher accuracy. Our technique easily generates several variants of the base architecture. Each variant returns only $\boldsymbol{2k}$ outputs $\boldsymbol{1 \leq k \leq \frac{\#classes}{2}}$, representing $\boldsymbol{Top\mhyphen k}$ classes, instead of tons of floating point values required in \modelp{}. Since each variant provides a full-class prediction, our approach maintains higher availability compared with \modelp{}  and \classp{}  in presence of failure.
\end{abstract}

\vspace{-.25em}

\vspace{-1.5em}
\section{Introduction}
\label{sec:introduction}

%\textbf{Deep Learning and CNN:}
\IEEEPARstart{C}{onvolutional} neural networks (CNNs) 
are being used in several visual analysis tasks such as image recognition, 
object detection and segmentation \revise{1}{ \cite{liu2022convnet} and tasks such as video analytics \cite{bhardwaj2022ekya, khanirecl}.}
In CNNs, computation grows proportional to the input size, and it eventually leads to higher latency. 
Although reducing the number of parameters and precision of variables help in improving the latency, 
they often come at the cost of lower accuracy \revise{1}{\cite{sandler2018mobilenetv2,tan2021efficientnetv2}.}
Moreover, the complexity of deep learning models and their applications are only to grow over foreseeable future, and hence, overall they require even more resources. 
{Model distribution across multiple machines is one viable solution that has consequently attracted growing attention especially when considering the fact that there are typically quite a number of local compute nodes around us that can help get the jobs done.}

\begin{figure} 
  \centering
\subfloat[\footnotesize Model Parallelism\label{fig:mp_architecture}]{%
      \includegraphics[width=0.25\textwidth]{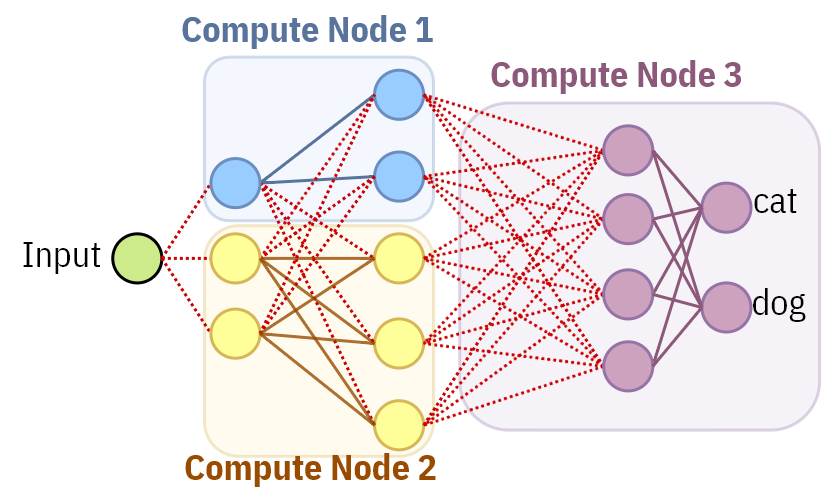}}
	~
%   \hfill
\subfloat[\footnotesize Class Parallelism\label{fig:cp_architecture}]{%
      \includegraphics[width=0.25\textwidth]{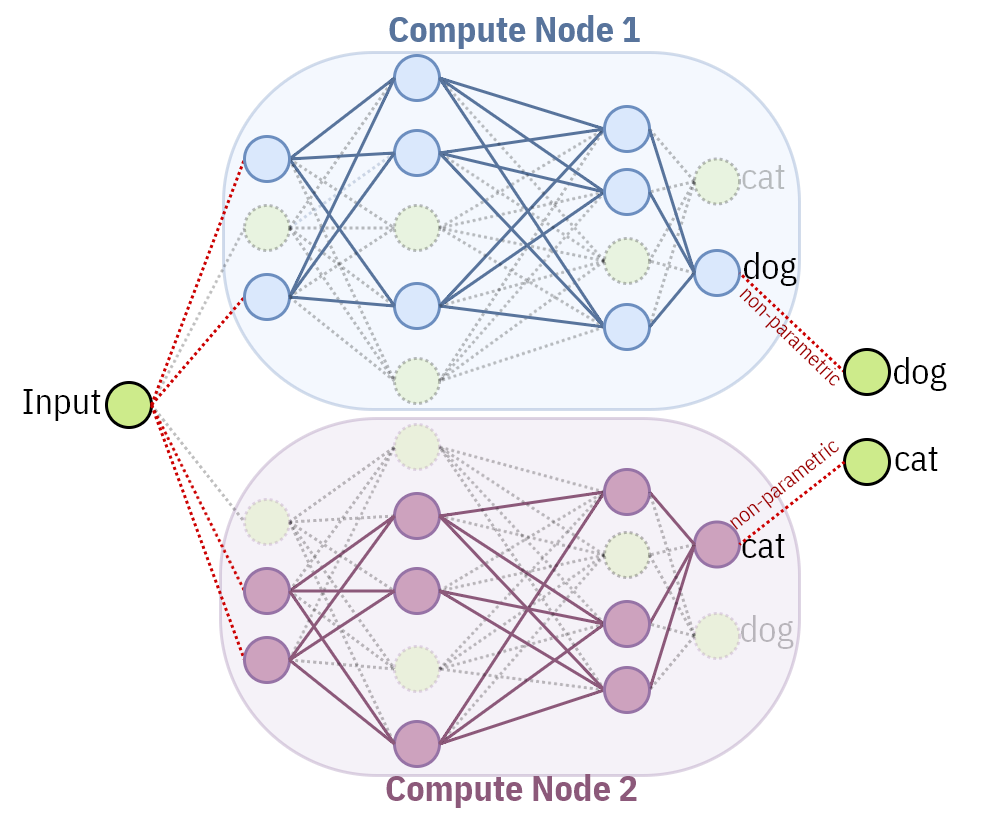}}
%   \hfill
	% ~
	\\
\subfloat[\footnotesize Variant Parallelism\label{fig:vp_architecture}]{%
      \includegraphics[width=0.25\textwidth]{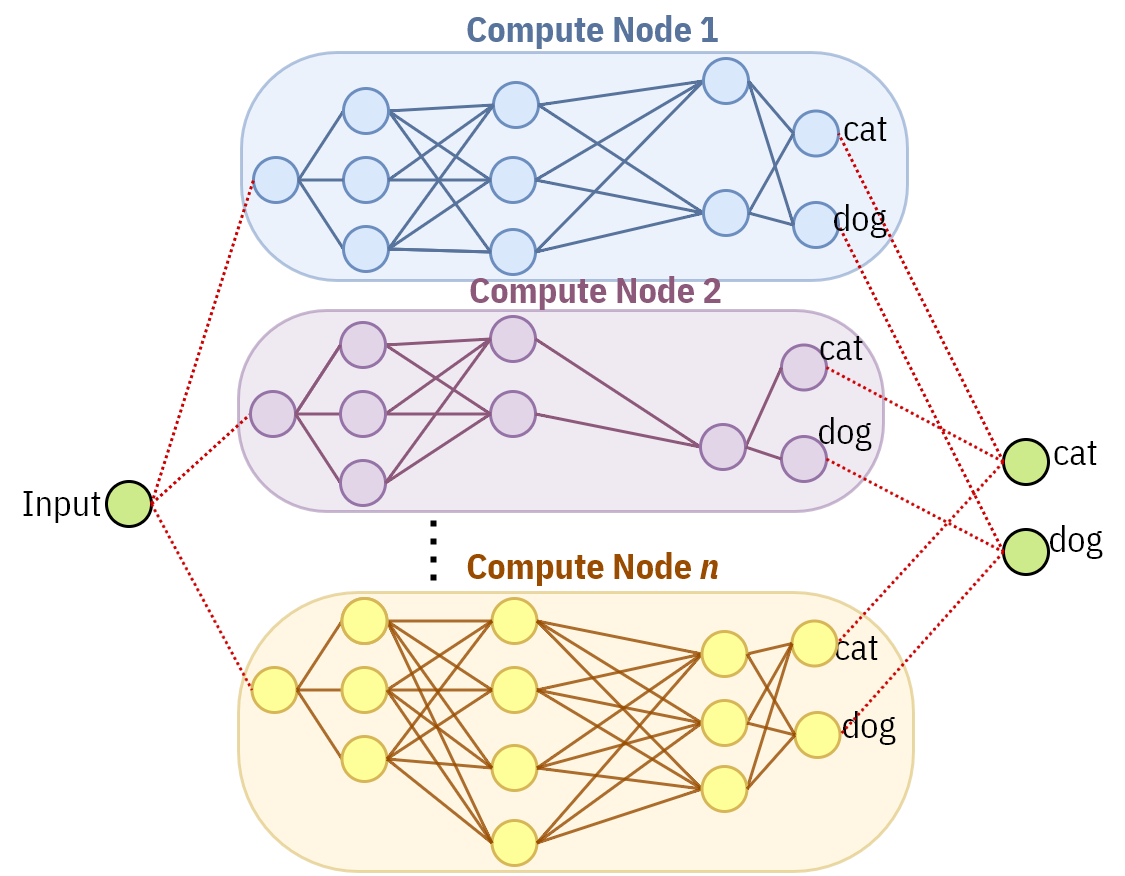}}
\caption{Comparison of different model distribution methods on atomic input data. \revise{1}{In (b), some neurons and connections (opaqued in the picture) are pruned so that each model predicts only one class to improve overall performance.} }
\vspace{-2em}
\label{fig:mp_cp_cp_aio} 
\end{figure}

For atomic data that cannot be split into further pieces \cite{MLSYS2021_sensai}, two major solutions are often used to decompose and distribute a model accross compute resources; namely, model parallelism (\modelp{}) and class parallelism (\classp{}).
\modelp{}  decomposes a model into multiple sequential or parallel slices, each sent over to a separate compute node for processing, then lots of intermediate data values are returned to the master to combine and complete the model. 
Slicing in \modelp{}  can be inter- or intra-layer (Fig.~\ref{fig:mp_architecture}).
\classp{}  decomposes the model into multiple disjoint models, each aiming to predict only one or a few non-overlapping classes (Fig.~\ref{fig:cp_architecture});
thus, \revise{1}{contrary to \modelp{}, \classp{} runs a monolithic model on each computing node.}
\classp{} improves efficiency of each model mostly by applying class-aware pruning policies \cite{MLSYS2021_sensai}.

\revise{1}{Both solutions have their own limitations. In \modelp{}, intermediate tensors have to be communicated between different nodes.}
%As illustrated in XXXFIGXXX, in 
In current CNNs, the time needed to transmit intermediate tensors can be orders of magnitude higher than inference time itself.
%the amount of data required to be transmitted is usually one to two orders of magnitude larger than the input and output of network itself. 
Hence, \modelp{}  would drastically increase end-to-end latency \cite{emmons2019cracking}. 
\classp{}  does not suffer from this issue since there is no sequential dependency and each computing node only returns a few numbers; 
these numbers represent the prediction of classes allocated to the submodel executed on that node. 
Verbatim \classp{}  has a firm restriction on scalability that is limited to the number of classes in the application \cite{MLSYS2021_sensai}, which can debatably be extended by choosing overlapping subsets of classes instead.
\revise{1}{However more importantly, both paradigms intrinsically suffer from multiple instances of single point of failure (SPOF) issue because every compute node has to complete its job and send the result back to the requesting node before the top-level classification task can be fully finished. 
Otherwise, the master node cannot generate the final network prediction.}
Therefore, failure in a single node compromises availability of the whole system. 
The scenario of interest in this paper is where we use the free times of IoT devices to do a distributed inference job.
Failure is a common problem in distributed systems, especially in this scenario of cycle-stealing from nearby IoT devices, where worker nodes only contribute in idle times and at any times may preempt the inference task to process their main jobs \cite{tan2023stateos}, and some nodes may seriously slow down due to various unexpected reasons. Additionally, communication issues such as network congestion, lead to missed deadlines in real-time applications. 
These concerns motivate the need for a fault-tolerant mechanism.
Obviously, in both \modelp{}  and \classp{}, higher availability may be achieved by introducing redundant nodes, but of course this comes at the expense of large impact on communication and computation cost.

%\textbf{Variant Parallelism; Its advantages:}
%\textbf{Intuition behind Variant Parallelism:}
In this paper, we propose a distribution technique: Variant Parallelism (\variantp{}).
Both \modelp{}  and \classp{}  can be viewed as applying a ``top-down" model slicing approach: they try to break a baseline model into pieces, although differently in \modelp{}  vs \classp{}, and distribute each piece across multiple nodes. 
In contrast, we in \variantp{} employ a ``bottom-up" approach:  we generate multiple variants (described below) from a chosen basic architecture
%  (e.g. MobileNetV2 \cite{sandler2018mobilenetv2}),
  and then combine their results in an aggregator component (Fig.~\ref{fig:vp_architecture}). 
In fact, \classp{}  and \variantp{} are both special types of ensemble learning models, but with different intents and approaches.

\revise{1}{In \variantp{}, each variant is an independent model that can be deployed on a different machine or accelerator. Unlike \classp{}, our variants have full-class prediction heads, i.e., each worker node predicts probabilities for all classes, and this is the key in achieving a fault-tolerant architecture. 
Consequently, in a system with $n$ worker nodes, \variantp{} can still get a full response even if $n\mhyphen 1$ nodes fail, while having the variants helps to improve accuracy over na\"ive replication when more than one worker node is working.}
\variantp{}'s variants can differ in input resolution and input offset. 
To achieve faster inference time, we apply three network optimizations: narrowing network width, reducing complexity of the compute intensive layers, and replacing with faster operations when possible.

Since each machine generates full-class prediction, the data transmission amount would be higher than \classp{}, resulting in risking a higher response time. 
To avoid this, we put a compression/decompression module. The compression layer attaches to each worker's classification head. 
In the compression part we select $k$ classes with the highest confidence score and send them to the master together with their corresponding indices. 
% In decompression part in the master, we reproduce full-class prediction result before aggregation. The intuition behind this heuristic is that except for a few classes with high scores, others are close to zero, which would have even added noises that negatively impact the aggregated results.
In \variantp{}, each machine executes a lightweight classification model, and sends its compressed result to the master. The master machine then applies a decompression step followed by a score scaling to weight each prediction based on its corresponding model capacity. 
It eventually combines results via an aggregation module. 
\revise{1}{There is no trainable parameter in the combination modules. Otherwise, it reduces system's flexibility and can compromise availability. }
\revise{1}{For evaluation, we consider a smart home scenario and demonstrate \variantp{}'s robustness on seven object recognition benchmarks. }
% Since our ultimate goal is to provide a flexible network architecture for users of smart home/environment scenarios, we chose those datasets that can be trained on a mediocre machine with a single GPU. However, the idea can be easily expanded to more complicated tasks.

Our major contributions can be summarized as follows:
\begin{itemize}
\item We introduce a bottom-up model distribution called Variant Parallelism which contrary to \modelp{}  and \classp{}  is fault-tolerant, flexible, and therefore, can be freely distributed across several sporadically available compute nodes such as the case of IoT devices or across multiple accelerators within a single machine.
\item Our technique generates multiple variants each having different number of parameters, multiply-accumulate operations (MACs), latency, and model size that can be deployed on different machines based on their capacity.
\item We propose an aggregation method that is resilient to failures, while matching accuracy of the baseline.
\item We propose a fast and simple technique to reduce bandwidth usage. It can compress the output size of each node by up to $10\times$ while losing less than $0.5\%$ accuracy. \revise{1}{The gain would be two orders of magnitude higher on benchmarks with large number of classes, e.g., ImageNet.}
\item Our proposed method can have $5.8\mhyphen 7.1\times$ fewer parameters, ${4.3\mhyphen 31\times}$ fewer MACs, and ${2.5\mhyphen 13.2\times}$ lower response time on an atomic input compared to the baseline while achieving comparable or higher accuracy.
% \item We get the same accuracy while requiring fewer machines, and having lower response time compared with the current state-of-the-art parallelism algorithms.
% \item We make our code publicly available \footnote{\url{github.com/Scalable-DNN/variant-parallelism}}.
\end{itemize}

%====================================================================================================

\section{Related Work}	
\label{sec:related_work}

\textbf{Model Parallelism (\modelp{}):} In model parallelism (Fig.~\ref{fig:mp_architecture}) a deep learning model is sliced into multiple sub-models \revise{1}{\cite{zeng2020coedge,DBLP:conf/mlsys/YuC20}}. Each of them can be deployed and concurrently and/or sequentially executed on multiple machines. These sub-models have to transfer data from intermediate tensors. The time to transfer these tensors can be orders of magnitude higher than the time to compute operations of the neural network itself. \revise{1}{For example, \cite{emmons2019cracking} shows that the intermediate outputs of deep learning models are $19$-$4500{\times}$ larger than the compressed input.} 
\revise{1}{Recent proposals' focus has mostly been on reducing communication rounds \cite{du2020distributed} or its overhead \cite{zhou2022accelerating}. Though, the SPOF problem is still inevitable and inherent to MP.
}

\textbf{Class Parallelism (\classp{}):} It decouples a base model into multiple independent sub-models, each can predict one or a few non-overlapping classes (Fig.~\ref{fig:cp_architecture}). Structured pruning is then applied on each sub-model to reduce their latency. The ultimate result can be prepared by combining the prediction of all of these sub-models \cite{MLSYS2021_sensai,yang2020robust}. 
\classp{}  does not have the issue mentioned for \modelp{}, and can improve latency compared to a single model. However, it has a hard limitation on the number of sub-models. More importantly, both \modelp{}  and \classp{}  require results from all machines and suffer from mutiple SPOFs.

% \textbf{Data Parallelism (\emph{DP}):}
% There is also another parallelism paradigm called data parallelism, in which all machines run the same model, but the data that they observe is different \cite{MLSYS2019_c74d97b0,DBLP:conf/mlsys/WangVPTDS20}. For example, each may process a different part of a single image. However, for atomic live data which is impossible to be further split, \emph{DP} cannot be leveraged.

Variant parallelism (\variantp{}) works on atomic live data and by design tolerates up to $n\mhyphen 1$ nodes failure. Contrary to other techniques, \variantp{} gives us the \textit{flexibility} to generate enough variants based on different objectives including latency constraints, \revise{1}{model size (proportional to the number of parameters)}, and compute resources. \variantp{} can be easily combined with other parallelism schemes.

%====================================================================================================

\section{Variant Parallelism}
\label{sec:variant_parallelism}

\begin{figure}[!t]
    \centering
    \includegraphics[width=0.8\columnwidth]{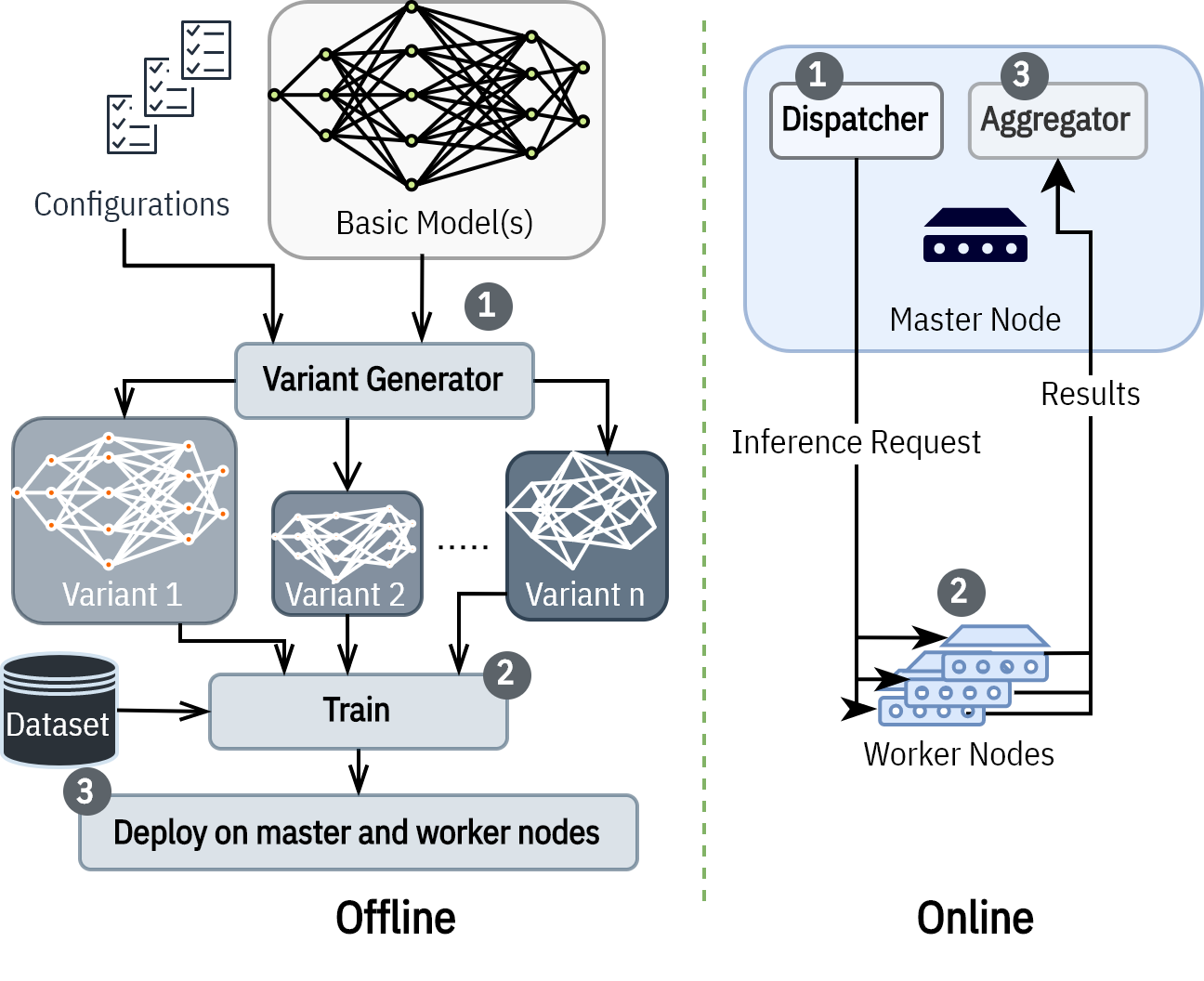}
	\vskip -0.15in
	% \caption{An overview of different steps in generation, training and dispaching modules.}
    \caption{Overview of our proposed workflow including offline parts (variant generation and model training) and online actions (dispatcher and aggregator).}
    \label{fig:workflow}
	% \vskip -0.2in
	\vspace{-1em}
\end{figure}

In \variantp{}, we consider a basic model, and generate multiple lightweight and fast models based on it. 
Each of these variants can have different storage sizes, parameters, inference times, and accuracies. 
% TODO: explain stochastic features : DONE
We add inference-time augmentation, \revise{1}{that leads to feeding slightly different inputs for each model.}
\revise{1}{Each variant can be treated as an independent full-class predictor: Every machine can concurrently execute one or more of these variants in-parallel with other machines. }
Depending on the use-case, these machines may be different GPUs on a single node or different edge devices, which is the case of interest in this paper. 
If there is an idle node with limited storage space or computation constraints, \variantp{} can generate a lighter variant which provides the opportunity to utilize the available compute capacity. 
\variantp{} aims to reduce response time to enable faster decision making via a distributed deep learning architecture while maintaining availability in presence of failure. 
% Here, we first concisely overview \variantp{}'s workflow. We then explain our design decisions, and how each module works.
\vspace{-1em}
\subsection{Overview}
\label{sec:vp,sebsec:overview}

\begin{figure}[!t]
    \centering
    \includegraphics[width=.85\columnwidth]{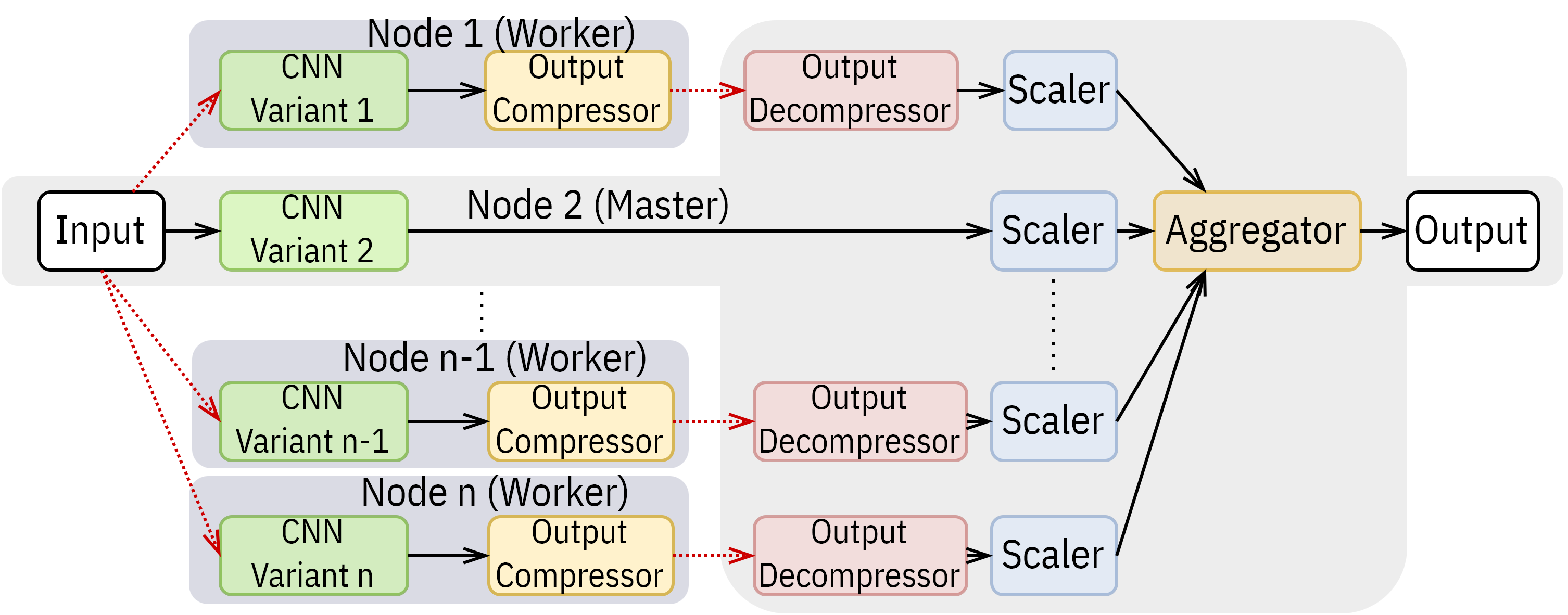}
	\vskip -0.12in
    \caption{The flow of data and the online operations on them, for inference.}
    \label{fig:system_overview}
	\vspace{-1.5em}
\end{figure}

Our workflow starts with one basic architecture, but is straightforward to extend to more than one model. 
% Here, we choose the efficient MobileNetV2. 
As shown in Fig.~\ref{fig:workflow}, we generate different variants from the base model(s). 
We then add classification head, compression, decompression, and scaling modules to each variant.
Variants are retrained on the desired dataset. 
These variants are independent and can be separately deployed on different machines for distributed in-parallel execution. 
Every node has its own model. Once deployed, the master node multicasts live data to all contributing workers, and waits for a predefined period (proportional to the desired deadline).
The workers feed the data into their models and send the compressed result back to the master node. 
The master then decompresses, scales and combines them through the aggregation module to prepare the final prediction (Fig.~\ref{fig:system_overview}).

\vspace{-1em}
\subsection{Reducing Model Complexity}
\label{sec:vp,subsec:optimization}

For variant generation, we reduce complexity of the basic architecture through the following three steps: (1) reducing width factor, (2) reducing complexity of the last few layers, and (3) replacement with faster operations when possible.
These steps help us to improve per machine model size, computation requirements, and latency.
Firstly, we reduce width factor of depth-wise separable convolution layers (was first introduced in \cite{howard2017mobilenets}) by $2.85{\times}$ ($2{\times}$ for ImageNet variants). 
It improves network runtime \revise{1}{\cite{howard2017mobilenets}}.
We recover the lost accuracy by combining multiple parallel variants (\S\,\ref{sec:experiments,subsec:scaling_acc}).
Secondly, we reduce complexity of the last few layers of each variant based on its input size. 
The intuition is that in the current efficient CNNs, the last few layers take considerable portion of the computation time. 
However, it should be proportional to the input resolution and the achievable accuracy. 
As shown in Table~\ref{table:models_architectures}, each model uses convolutional layers with the number of filters ($f$) proportional to its input size ($\rho_i$). 
For example, input size of $V_3$ ($i{=}3$) is $160$ and its last convolution layer has $384$ output filters while $V_1$ ($i{=}1$) with input size of $96$ has only $256$ output filters. 
We empirically select the number of these output filters so that they are divisible by $64$ for better performance.
\revise{2}{In that sense, the first and second steps are types of structured pruning that result in fewer parameters and MACs.}
Lastly, we employ faster operations when possible, e.g., replacing $FullyConnected$, $AveragePooling$, and $Average$ layers by $1\times1 Convolution$, $MaxPooling$, and $Add$ layers, respectively. 
We generate \revise{1}{five (for CIFAR-10, CIFAR-100, MNIST, F-MNIST and SVHN) and seven (for ImageNet and Food101)} different models based on these optimizations. 
Their characteristics are illustrated in Table~\ref{table:models_characteristics}.

\begin{table}[b]
	\vskip -0.2in
	\caption{Variants model architecture.}
%	\vskip 0.05in
	\vspace{-1em}
	\label{table:models_architectures}
	\begin{center} \begin{small} \begin{tabular}{p{2.5cm}llccc}
		% \hline
    \hline
		Input									& Operator			& $t$	& $f$				& $n$	\\
		\hline
		$3\times\rho_i^2$							& conv2d			& -		& 32				& 1		\\
		$32\times\rho_i^2/4$				& bottleneck		& 1		& 16				& 1		\\
		$16\times\rho_i^2/4$				& bottleneck		& 6		& 24				& 2		\\
		$24\times\rho_i^2/8$				& bottleneck		& 6		& 32				& 3		\\
		$32\times\rho_i^2/16$				& bottleneck		& 6		& 64				& 4		\\
		$64\times\rho_i^2/32$				& bottleneck		& 6		& 96				& 3		\\
		$96\times\rho_i^2/32$				& bottleneck		& 6		& 160				& 3		\\
		$160\times\rho_i^2/64$				& bottleneck		& 6		& 320				& 1		\\
		$320\times\rho_i^2/64$				& conv2d $1\times1$	& -		& ${64{\times}j}$	& 1		\\
		$(64\times j)\times\rho_i^2/64$	& globalmaxpool		& -		& -					& 1		\\
		$(64\times j)$			& conv2d $1\times1$	& -		& $C$				& -		\\
		$C$									& compression 		& -		& $k$				& -		\\
    \hline
		% \hline
    % \multicolumn{4}{l}{Similar to the original MobileNetV2, we apply a convolution layer on input followed by $19$ depth-wise convolutional residual bottlenecks. Each row is sequentially applied $n$ times, and $t$ is the expansion factor of intermediate layers in bottleneck blocks. $C$ is the number of classes.}
		\end{tabular} \end{small} \end{center}
    {\raggedright \footnotesize Similar to the original MobileNetV2, we apply a convolution layer on input followed by $19$ depth-wise convolutional residual bottlenecks. Each row is sequentially applied $n$ times, and $t$ is the expansion factor of intermediate layers in bottleneck blocks. $C$ is the number of classes. \revise{1}{$j{=}(3+i)$ when width-factor=0.35 and $j{=}\min\left[(7+2i),20\right]$ when width-factor=0.5 (for ImageNet variants)}.\par}
	% \vskip -1.3in
	% \vspace{-.5em}
\end{table}

\begin{table}[b]
	% \vskip -0.2in
	\caption{\revise{1}{Comparison of different models' characteristics}}
	% \vskip -0.5in
	\label{table:models_characteristics}
%	\vskip 0.05in
	\vspace{-1.5em}
	\begin{center} \begin{small} \begin{tabular}{p{1.5cm}p{1.3cm}p{3.4cm}p{.7cm}}
		\hline
		Model				& Input Size		& \#Params	& \#MACs	\\
		\hline
		MobileNetV2			& $3\times224^2$	& 2.27M (3.47 on ImageNet)		& 300M		\\
		$V_7$				& $3\times320^2$	& 410K		& 112M		\\
		$V_6$				& $3\times256^2$	& 390k		& 71M		\\
		$V_5$				& $3\times224^2$	& 370K		& 54M		\\
		$V_4$				& $3\times192^2$	& 360K		& 40M		\\
		$V_3$				& $3\times160^2$	& 340K		& 27M		\\
		$V_2$				& $3\times128^2$	& 330K		& 17M		\\
		$V_1$				& $3\times96^2$		& 320K		& 10M		\\
		\hline
		$V_{im7}$			& $3\times320^2$	& 1.97M		& 196M		\\
		$V_{im6}$			& $3\times256^2$	& 1.9M		& 125M		\\
		$V_{im5}$			& $3\times224^2$	& 1.75M		& 95M		\\
		$V_{im4}$			& $3\times192^2$	& 1.59M		& 69M		\\
		$V_{im3}$			& $3\times160^2$	& 1.45M		& 47M		\\
		$V_{im2}$			& $3\times128^2$	& 1.3M		& 30M		\\
		$V_{im1}$			& $3\times96^2$		& 1.15M		& 17M		\\
		\hline
		\end{tabular} \end{small} \end{center}
	% \vskip -0.3in
	{\raggedright \footnotesize \revise{1}{Variants $V_{im[i]}$ are exclusively designed for ImageNet benchmark \cite{deng2009imagenet}.}\par}
\end{table}

\vspace{-1em}
\subsection{Compression and Decompression}
\label{sec:vp,subsec:compress_decompress}

In \classp{}, for tasks with only few classes (e.g., CIFAR-10), each machine returns only one floating point, and for tasks with more classes (e.g., CIFAR-100), 10 to 20 floating point values are returned by each. 
In contrast, our variants are independent models that can return full-class prediction. This feature increases the system reliability in a faulty environment, but raises a challenge in communication time. 
To address the mentioned issue, we leverage the $Top\mhyphen k$ idea in which from a prediction vector $\psi_i$, we select $k$ elements having the highest confidence scores and pack them along with their corresponding indices. 
Fig.~\ref{fig:comp_decomp_scaling} depicts a toy example with 6 classes, assuming $k=2$. 
The compression step shrinks the vector size to two vectors of size $k$: One stores floating point predictions with the highest score ($0.51$ and $0.4$), and the second keeps their integer indices ($1$ and $3$). 
After receiving compressed vectors, the master reconstructs a tensor of shape ($batch\_size, V, C$), where $V$ and $C$ are the number of total contributing machines and classes, respectively. 
It then scatters values of compressed vectors based on their indices. This procedure is fast yet robust. 
The intuition behind is that in each vector except a few elements with high confidence scores, others are close to zero. 
More importantly, our aggregation method (\S~\ref{sec:vp,subsec:aggregation}) amplifies both actual information and noise at the same time. 
Hence, by removing values with lower confidence score, we actually mitigate noisy information to some degrees, and consequently achieve higher accuracy.
% The $Top\mhyphen k$ technique also gives us the flexibility to set $k$ as a hyperparameter, trading accuracy for transmission time. 
% In general, we can transmit even fewer amount of data than \classp{}  while maintaining the system reliability at a high level.

\vspace{-1em}
\subsection{Scaling and Aggregation}
\label{sec:vp,subsec:aggregation}

\begin{figure}[t!]
	\centering
	\includegraphics[width=0.8\columnwidth]{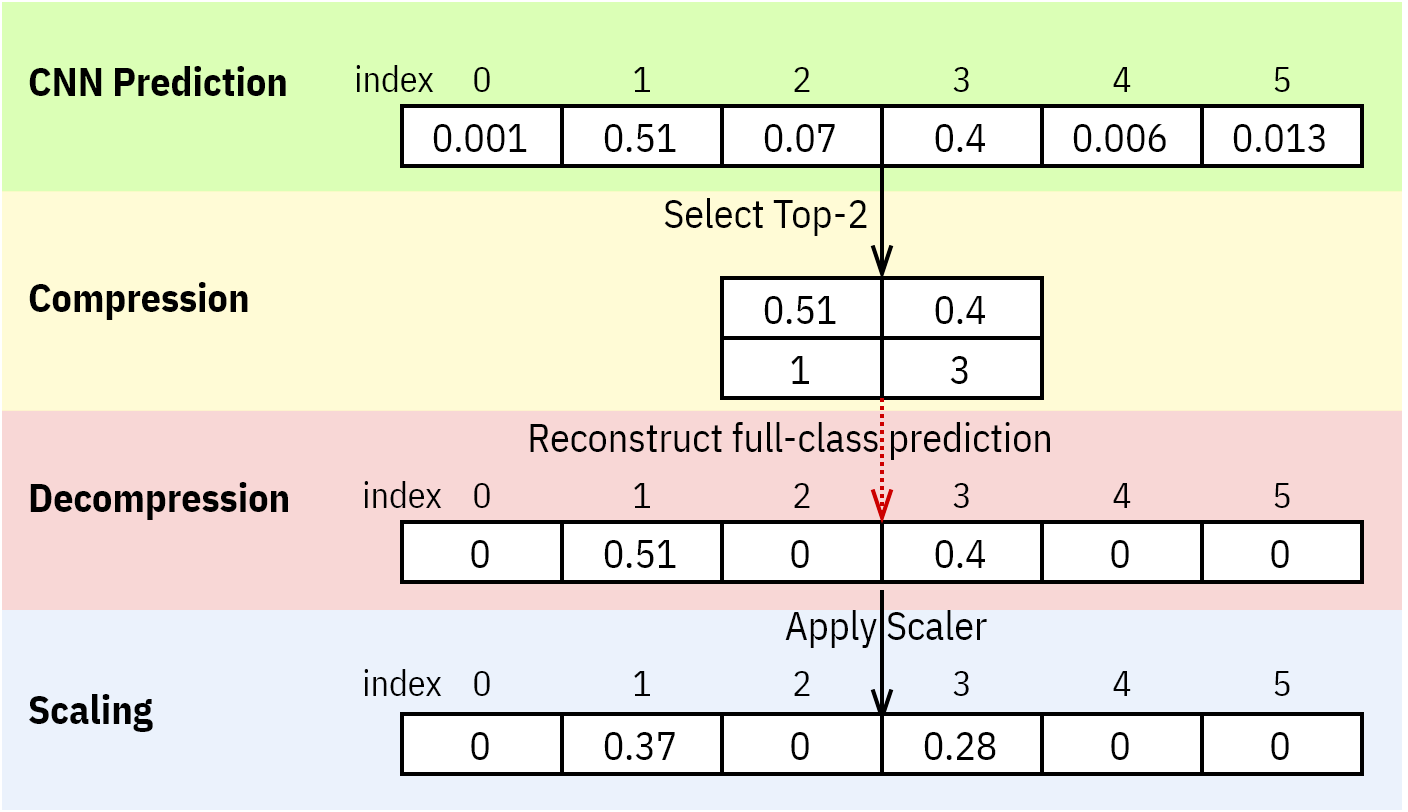}
	\vskip -0.15in
	\caption{An example of compression, decompression and scaling steps assuming $C=6$ and $k~(in\:Top\mhyphen k)=2$.}
	\label{fig:comp_decomp_scaling}
	% \vskip -0.2in
	\vspace{-1.5em}
\end{figure}

Master is responsible for balancing the prediction vectors of different variants. 
After decompression step, we apply a $Softmax$ operation to transform the reconstructed prediction vectors into a probability distribution (Fig.~\ref{fig:comp_decomp_scaling}).
We then need to scale each prediction vector based on its network capacity.
For example, if variant $V_i$ has more parameters, and thereby returns more accurate predictions than variant $V_j$, then it deserves to have a higher weight in the aggregated result. 
% An idea is to give each variant a weight proportionate to its single validation accuracy. Nonetheless, we may not have their results from the beginning. 
% In addition, a variant's accuracy is subjected to change as data distribution alters in different datasets. 
We use a proxy of the network architecture itself to scale values of every prediction vector via eq.~\ref{scaling_eq}.

\begin{equation}
\label{scaling_eq}
\psi_i = (\frac{\rho_i}{\max(P)})(\frac{d_i}{\min(D)})^\alpha.\psi_i
\end{equation}

\begin{equation}
where ~~ \rho_i \in P ~~, and ~~ d_i \in D
\end{equation}

Where $\psi_i$ is the prediction vector of variant $V_i$ with input shape $\rho_i$ and depth multiplier $d_i$, and $\alpha$ is a parameter to equalize the impact of input shapes and depth multiplier.

\revise{1}{In the aggregation module we combine prediction results that are received from compute nodes including the master itself.}
% Since we train multiple models together, an idea that may quickly come in mind is to concatenate the prediction vectors and apply one or more fully-connected ($fc$) layers. 
% Since these $fc$ layers have trainable parameters, we can get higher accuracy. However, employment of an aggregation module with trainable parameters has two drawbacks. 
% First and foremost, it indirectly damages the availability of the whole system. Our objective is to get a response even if all but one of worker nodes fail. 
% An $fc$ layer requires all of its inputs before processing any data, but one or a few of them may not be available. 
% Second, addition of one or more of these layers increases the total computation cost. 
\revise{1}{We add the vectors together and apply a $Softmax$ afterwards.
Using FullyConnected ($fc$) layers instead, although might improve accuracy, degrades flexibility. 
Furthermore, with the same number of neurons as the number of classes $C$, and $V$ variants contributing in the aggregation step, 
the computation cost of using only a single $fc$ layer would be $O(CV)$ which is $O(V)$ times higher.}
% It is fast yet amplifies confidence score of those classes with relatively higher probability in multiple vectors. 

% \subsection{Sources of Variation}
% \label{sec:vp,subsec:aggregation}

\section{Experiments}
\label{sec:experiments}

\subsection{Experimental Setup}
\label{sec:experiments,subsec:setup}

\noindent\textbf{Compute Platform:} 
We use an in-house server (Table~\ref{table:server_spec}), to design, train and evaluate our models. 
% This server has only $32GB$ main memory, but since multiple people share the server for different tasks, on average, $30\mhyphen70\%$ of its capacity was available for our experiments. 
% Because our ultimate goal is enabling end-users to train and deploy the models for their smart home applications, as previously mentioned, 
We selected the commonly reported vision datasets that are trainable on our server in a reasonable time. 

\begin{table}[!b]
	\vskip -0.2in
	\caption{Server specification}
	\vskip -0.2in
	\label{table:server_spec}
	\begin{center} \begin{small} \begin{tabular}{ll}
		\hline
		CPU		& Intel Xeon E5-2630-v3 (x86\_64)				 	\\
				& Frequency: 2.4GHz (1.2-3.2GHz)				   	\\
				% & Cache: 32K-32K-256K-20480K					   	\\
				% & \#Cores: 24 (double threaded)					   	\\
		GPU 	& NVidia GeForce GTX 1080 Ti					 	\\
			   	& Total Memory: 11GB							   	\\
				% & CUDA V9.0.176 , Compute Capacity 6.1							   	\\
		Memory	& 32GB (1600MHz)								   	\\
		% OS		& Gnu/Linux Ubuntu 18.04							\\
		\hline
		\end{tabular} \end{small} \end{center}
	% \vskip -0.3in
\end{table}

\noindent\textbf{Benchmarks:} 
\revise{1}{We have provided six different mid-sized object recognition datasets that can be categorized into three complexity levels (in terms of the achievable accuracy by the baseline architecture): simple (CIFAR-10 \cite{cifar_paper}, Fashion MNIST (F-MNIST) \cite{xiao2017fashion}, Fashion MNIST (F-MNIST) \cite{xiao2017fashion}, MNIST \cite{lecun2010mnist} and SVHN \cite{netzer2011reading}), 
mediocre (CIFAR-100 \cite{cifar_paper}) 
and difficult (Food-101 \cite{bossard2014food}). 
For completeness, we additionally provide accuracy  results on ImageNet-1k (i.e., 1k classes) \cite{deng2009imagenet}.
}

\begin{table}[b!]
	\vskip -0.2in
	\caption{\revise{1}{Datasets used for evaluation}}
	\vskip -0.2in
	\label{table:datasets}
%	\vskip 0.05in
	\begin{center} \begin{small} \begin{tabular}{p{1.5cm}p{1.2cm}p{1.5cm}cr}
		\hline
		Dataset     & $C$ & Input Size 	  & \#Trains & \#Tests \\
		\hline
		Food-101    & 101		& $3\times256^2 $ & 75,750 	 & 25,250  \\
		SVHN		& 10		& $3\times32^2 $  & 73,257	 & 26,032  \\
		CIFAR-10	& 10		& $3\times32^2 $  & 50k	 	 & 10k  \\
		CIFAR-100	& 100		& $3\times32^2 $  & 50k	 	 & 10k  \\
		MNIST		& 10		& $28^2 $  		  & 60k  	 & 10k  \\
		F-MNIST		& 10		& $28^2 $         & 60k	 	 & 10k  \\
		ImageNet    & 1000      & $3\times256^2$  & 1.28M	 	 & 50k \\
		\hline
		\end{tabular} \end{small} \end{center}
	% \vskip -0.3in
\end{table}

% \textbf{Graphic Card:} 
% We use an NVidia GeForce GTX 1080~Ti with $11GB$ Memory, which has rather good price to performance ratio.

\noindent\textbf{Edge Device:} 
For latency evaluation of different models, we use a mid-range tablet (Table~\ref{table:edge_spec}), and leverage Google TFLite native benchmark tool \cite{tflitebenchmark} to profile the performance.

\noindent\textbf{Baseline Model:} 
We consider MobileNetV2 \cite{sandler2018mobilenetv2} as our basic architecture (Table~\ref{table:models_characteristics}). \revise{1}{It is regarded as an efficient model designed for constrained edge devices. Thus, if we show gains in comparison to MobileNetV2, as we did, then one can expect similar results on models with more redundancy.}
% In our opinion, among different networks proposed in recent years, MobileNetV2 is more suitable for edge device constraints. 
% For example, EfficientNet-B0 \cite{Tan2019a} achieves better accuracy, but has twice as many parameters as MobileNetV2. 
% The characteristics of our baseline model is shown in Table~\ref{table:models_characteristics}.

\vspace{-1em}
\subsection{Training} \label{sec:experiments,subsec:training}

\revise{1}{We trained $V_1$ through $V_7$ for $135$ epochs. 
Our training procedure has two steps. For the first 30 epochs, all layers except the last two convolutions were frozen. 
For the ImageNet variants ($V_{im[i]}$), we iterate 75 epochs (the first 40 for fine-tuning).
We port pre-trained weights from ImageNet-1k \cite{deng2009imagenet} whenever available. 
We used batch size of 128 in this step, and 64 (32 in Food-101) for the rest of 105 epochs. We used Adam optimizer with $\beta_1=0.9$ and $\beta_2=0.999$. Learning rate decays by a factor of $0.98$ once the accuracy stagnates. We do not apply hyperparameter optimization. Thus, one may improve the results by doing so.
Since our tasks are classification, and each model can independently predict all of the classes, we apply a separate categorical cross entropy for each model. 
Contrary to \classp{}, our variants can observe the same data during training. 
Therefore, we train multiple models along with the combination components all together. 
The number of models that can be trained simultaneously is a parameter that can be set based on the computation capacity. 
Having more models requires more memory, but improves GPU utilization as most parts of our data pipeline executes once.
To avoid overfitting, we apply typicaly used augmentation methods namely random horizontal flip, random rotation and MixUp.}
%  \cite{zhang2017mixup} with $\alpha=0.2$. 

% \begin{figure} 
%   \centering
% \footnotesize \subfloat[\footnotesize First 30 epochs\label{fig:lr_sched1}]{%
%       \includegraphics[width=0.35\columnwidth]{lr_sched1}}
%   ~~~
% \subfloat[\footnotesize From epoch 30 until the end\label{fig:cp_architecture}]{%
%       \includegraphics[width=0.35\columnwidth]{lr_sched2}}
% \caption{Learning Rate Scheduling. Decaying was applied on plateau by a factor of 0.98.}
%   \label{fig:lr_sched}
% \end{figure}

\vspace{-1em}
\subsection{Accuracy of Single Models}	\label{sec:experiments,subsec:accuracy}

First, we evaluate accuracy of each variant and compare it to the accuracy of our baseline. 
To have a fair comparison, we trained the baseline with the same training policy, and made sure it has the same or higher accuracy compared with original reports, if applicable\footnote{\revise{1}{For ImageNet dataset, we could not reach the reported accuracy. This is likely due to the choice of hyperparameters and fewer iterations as it was not feasible for us to provide the same infrastructure (16 GPU machine vs. 1 GPU machine with lower computation capability). Nevertheless, since the same policy was applied to all models including the baseline, the results should hold fair.}}.
For each dataset, we report the accuracy of five (seven for Food-101 and ImageNet) different models. We also report $top\mhyphen 2$ and $top\mhyphen 5$ in aggregator for datasets with ${C<100}$ and ${C>=100}$, respectively. Here, by $top\mhyphen m$ we mean the ground-truth label is among top $m$ class predictions with the highest confidence scores. 
% Later, we show the use of multiple variants and our compression method not only increase $top\mhyphen 1$ accuracy, but also $top\mhyphen m$ accuracy. 
\revise{1}{As depicted in Table~\ref{table:single_models_acc}, as we go from the variant $V_1$ to $V_7$ ($V_{im1}$ to $V_im7$ on ImageNet), the accuracy improves. For the datasets with few number of classes, the accuracy of $V_5$ is close to the baseline. Nevertheless, our $V_5$ has ${\sim}6\times$ fewer parameters and MACs.}
\revise{1}{This shows that for simple tasks with few classes, we can already reach to the baseline accuracy level without the need for parallelism. In fact, our bottom-up ensemble design gives us the flexibility to pick only one or a few variants for a custom use-case scenario while in model and class parallelism, all models are mandatory. Yet, by using variant parallelism, as we explain next, we can achieve even higher accuracy levels compared with the baseline.}

\begin{table}[b]
	\raggedleft
	\vskip -0.2in
	\caption{Edge device specification}
	\vskip -0.2in
	\label{table:edge_spec}
%	\vskip 0.05in
	\begin{center} \begin{small} \begin{tabular}{p{1.2cm}p{6.2cm}}
		\hline
		\multicolumn{2}{c}{\textbf{Samsung Galaxy Tab A 8.0 with S Pen (2019, SM-P205)}} \\
		\hline
		CPU		& Octa-core big.LITTLE (2x1.8 GHz Cortex-A73 and 6x1.6 GHz Cortex-A53) 	\\
		% Chipset	& Exynos 7 Octa 7904 (14 nm) – 64bit				\\
		GPU 	& ARM Mali-G71 MP2								   	\\
		Memory	& 3GB LPDDR4X									   	\\
		% Storage	& 32GB (eMMC 5.1)									\\
		% OS		& Android 10										\\
		Network	& Wi-Fi 802.11, estimation: RTT=2ms~bw=1Mbps				\\
		\hline
		\end{tabular} \end{small} \end{center}
	% \vskip -0.32in
\end{table}

\begin{table*}[]
	\vskip -0.2in
	\caption{\revise{1}{Accuracy (\%) of single models on different datasets}}
	\vskip -0.2in
	\label{table:single_models_acc}
%	\vskip 0.05in
	\begin{center} \begin{small} \begin{tabular}{p{1.7cm}lcccccccc} 
		\hline
		Dataset		& Metric	& MobileNetV2	& $V_1$	& $V_2$	& $V_3$	& $V_4$	& $V_5$ 	& $V_6$	& $V_7$	\\
		\hline
		CIFAR-10	& Top-1		& 95.16			& 91.76		& 93.07		& 93.81		& 94.43		& 94.62	&-&-	\\
					& Top-5		& 98.57			& 97.24		& 97.87		& 98.20		& 98.26		& 98.29	&-&-		\\
		\hline
		MNIST		& Top-1		& 99.37			& 99.38		& 99.41		& 99.41		& 99.42		& 99.46	&-&-		\\
					& Top-2		& 99.87			& 99.92		& 99.95		& 99.89		& 99.92		& 99.95	&-&-		\\
		\hline
		CIFAR-100	& Top-1		& 80.51			& 69.62		& 73.57		& 74.46		& 74.67		& 75.20	&-&-		\\
					& Top-5		& 94.9			& 91.17		& 93.36		& 93.47		& 93.41		& 93.67	&-&-		\\
		\hline
		SVHN		& Top-1		& 95.52			& 93.81		& 95.16		& 95.34		& 95.89		& 95.91	&-&-		\\
					& Top-2		& 98.37			& 97.47		& 98.21		& 98.25		& 98.46		& 98.32	&-&-		\\
		\hline
		\multirow{2}{1.7cm}{Fashion-MNIST}
					& Top-1		& 94.41			& 93.38		& 94.16		& 94.60		& 94.67		& 94.55	&-&-		\\
					& Top-2		& 98.82			& 98.30		& 98.51		& 98.76		& 98.84		& 98.82	&-&-		\\
		\hline
		Food-101	& Top-1		& 79.68			& 58.78		& 65.35		& 70.13		& 71.43		& 74.23	& 74.64	& 75.79		\\
					& Top-5		& 93.5			& 83.50		& 87.70		& 90.19		& 90.78		& 92.43	& 92.52	& 93.15		\\
		\hline \hline
		Dataset		& Metric	& MobileNetV2	& $V_{im1}$ & $V_{im2}$	& $V_{im3}$	& $V_{im4}$	& $V_{im5}$ & $V_{im6}$	& $V_{im7}$	\\
		\hline
		ImageNet	& Top-1		& 67.5		& 48.8		& 54.7		& 57.9		& 61		& 62.25		& 63.2	& 64.9	\\
					& Top-5		& 87.9		& 73.65		& 78.7		& 80.9		& 83.3		& 84.2	& 85 & 86		\\
		\hline
		\end{tabular} \end{small} \end{center}

	% \vskip -0.15in
\end{table*}

\vspace{-1em}
\subsection{Impact of Compression Method}
\label{sec:experiments,subsec:compression}

In this section, we analyze the effect of our compression algorithm on accuracy and the required bandwidth.
We feed the same input data to all variants, and evaluate the impact of different values for $k$ in our compression module.
%, and select a specific value for the rest of the experiments.

% \subsubsection{Bandwidth Saving}	
\label{sec:experiments,subsec:aggregate_acc}
% \begin{figure}
% 	\centering
% 	\includegraphics[width=0.7\columnwidth]{k_experiment_output_size}
% 	\vskip -0.2in
% 	\caption{Effect of different values for $k$ in compression step on output size of a variant, assuming $\#classes=100$.}
% 	\label{fig:k_experiment_output_size}
% 	% \vskip -0.2in
% \end{figure}

% Fig.~\ref{fig:k_experiment_output_size} shows how different values for $k$ in our compression method impacts output size of a single variant.
\noindent\textbf{Bandwidth Saving:}
Output size is a linear function of $k$. 

\begin{equation}
	\label{k_outputsize}
	h(k) = 
	\begin{cases}
		\min \{ k\times[\text{size}(fp)+1], \text{size}(fp) \times C \} & k > 1 \\
		1 & k = 1
	\end{cases}
\end{equation}

Where $h(k)$ is the output size in byte, $\text{size}(fp)$ is the number of bytes required for storing a floating point value and $C$ is the number of classes.
Among different $k$ values, $k=1$ is a special case. Because the compressor selects only one element, we can further reduce communication cost, sending only an integer representing its index. 
% Depending on the number of classes, it can be transmitted as a single \textit{uint8} value 
% which leads to up to $10\times$ saving compared to compression with $k=2$. 
It can be found that choosing $k>\frac{C}{2}$ is not logical as transmitting the whole vector actually costs less than compressing it.
% This happens due to the overhead imposed to carry indices of each element.

% \subsubsection{Ensemble Accuracy}	
\label{sec:experiments,subsec:aggregate_acc}
\begin{figure*}[t!]
    \centering
    \subfloat[\footnotesize CIFAR-10 aggregator Top-1\label{fig:cifar10_top1}]{%
      \includegraphics[width=0.22\textwidth]{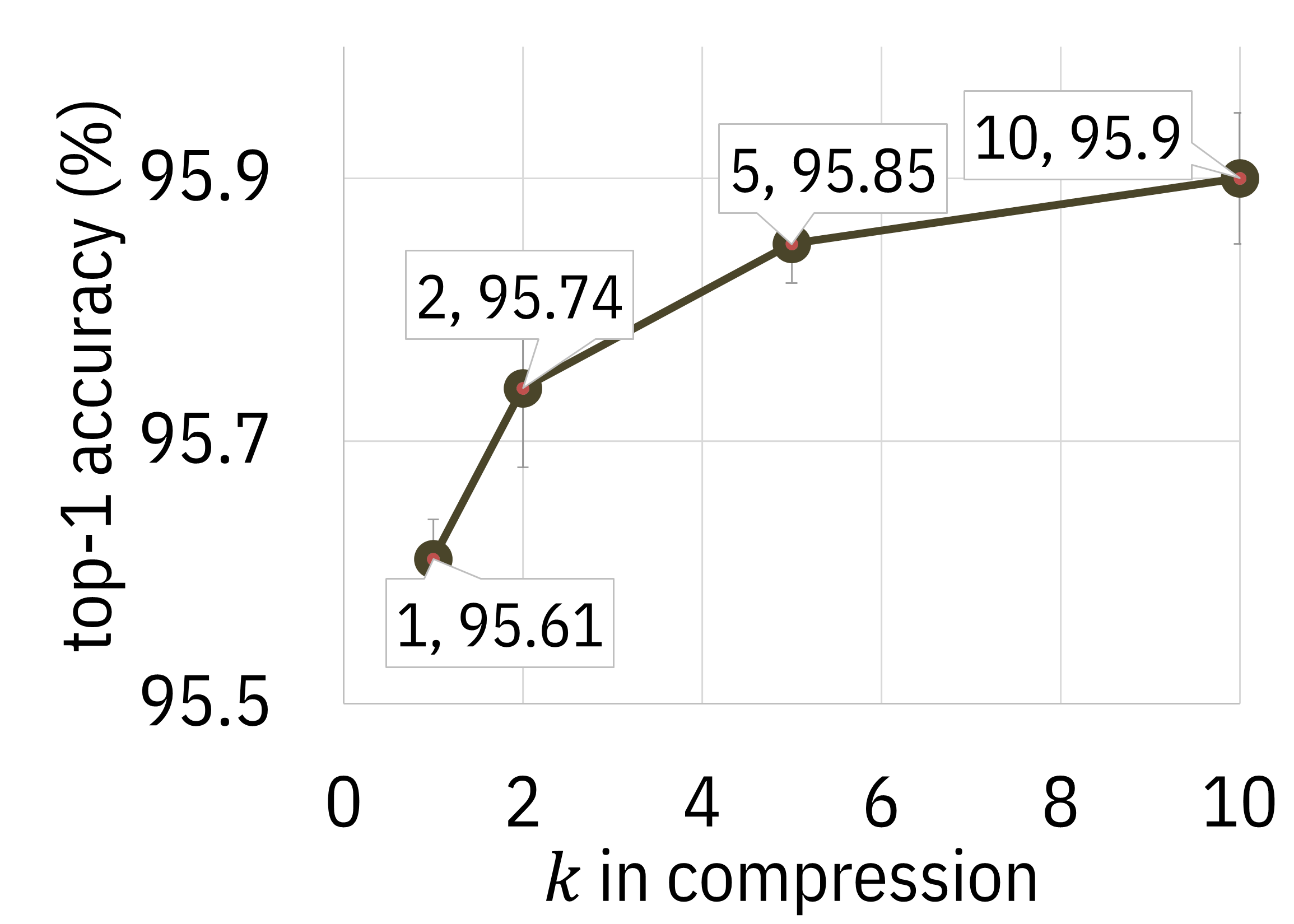}}
    ~
    \subfloat[\footnotesize CIFAR-100 aggregator Top-1\label{fig:cifar100_top1}]{%
      \includegraphics[width=0.23\textwidth]{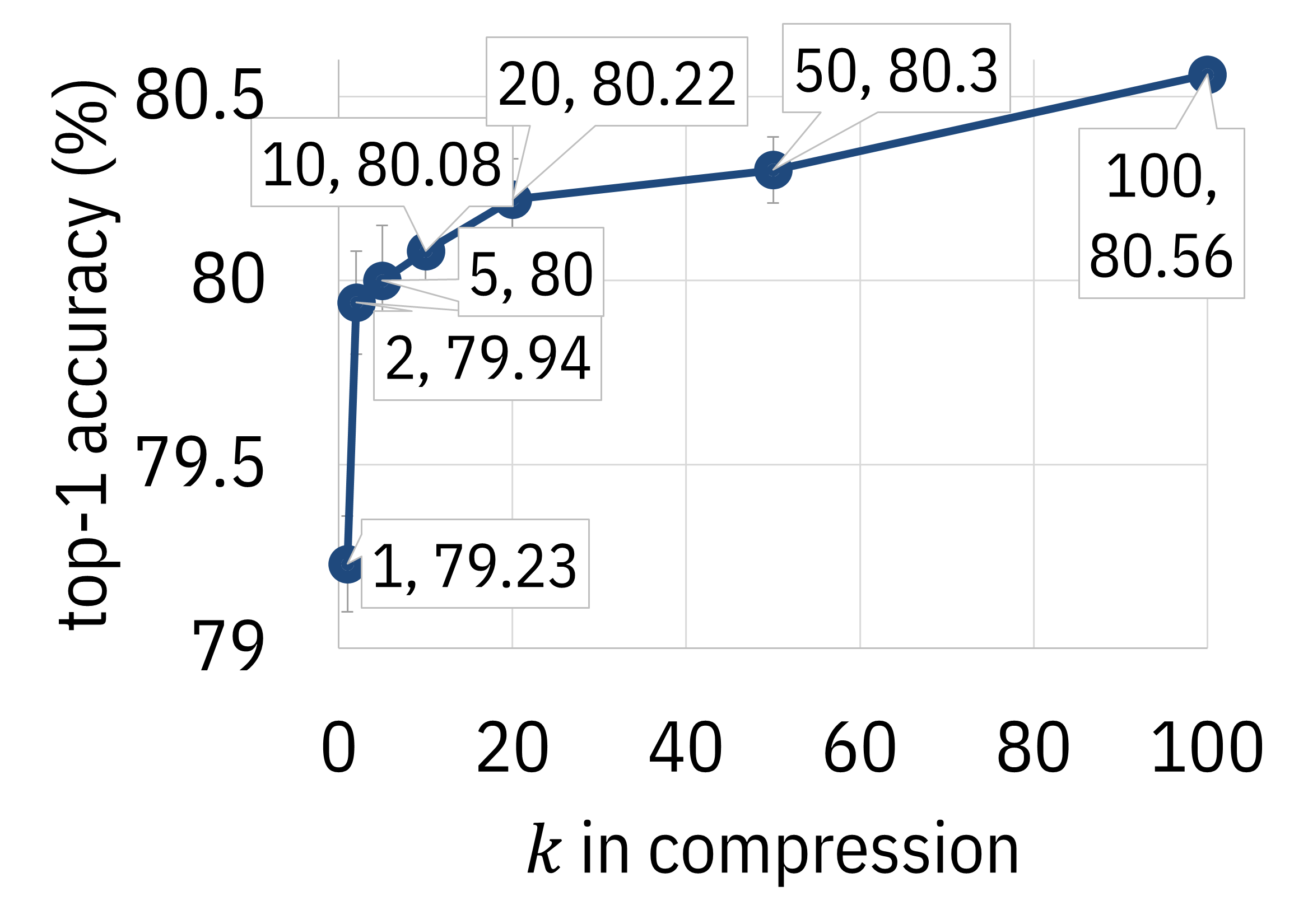}}
    ~
	\subfloat[\footnotesize Food-101 aggregator Top-1\label{fig:food101_top1}]{%
    \includegraphics[width=0.22\textwidth]{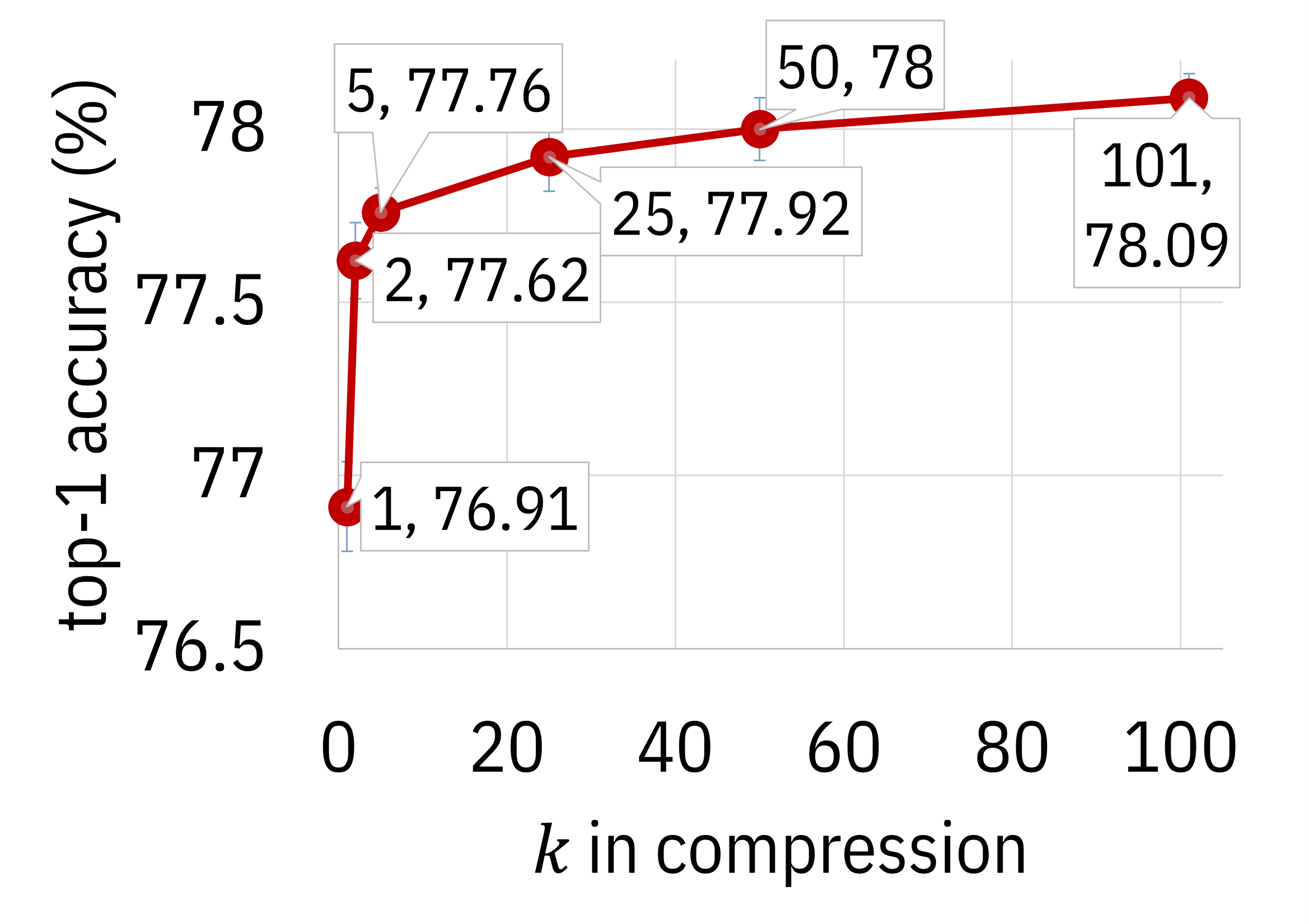}}
	~
	\subfloat[\footnotesize ImageNet aggregator Top-1\label{fig:imagenet_top1}]{%
    \includegraphics[width=0.22\textwidth]{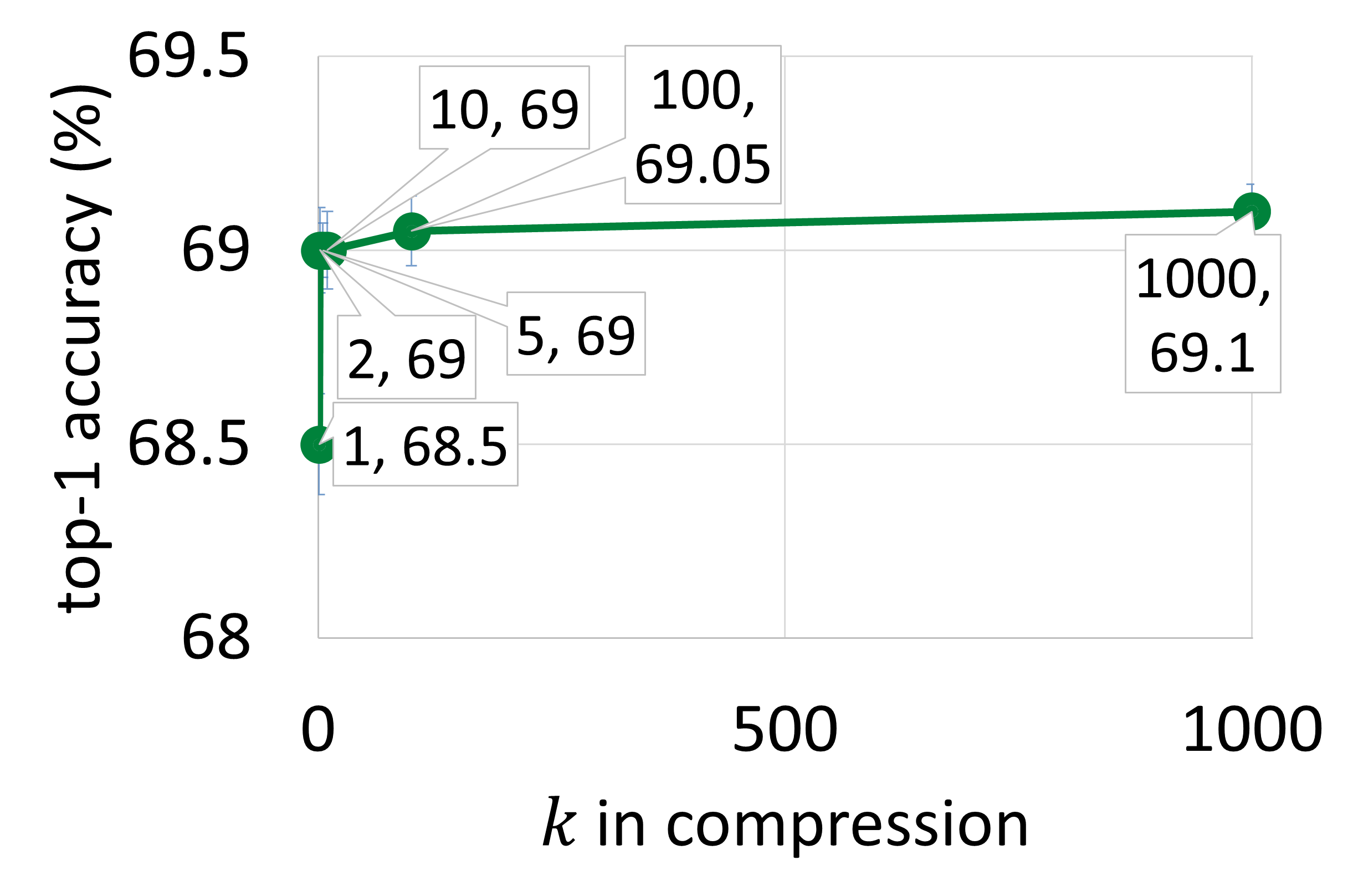}}
	\\
	\subfloat[\footnotesize CIFAR-10 aggregator Top-2\label{fig:cifar10_top2}]{%
	\includegraphics[width=0.22\textwidth]{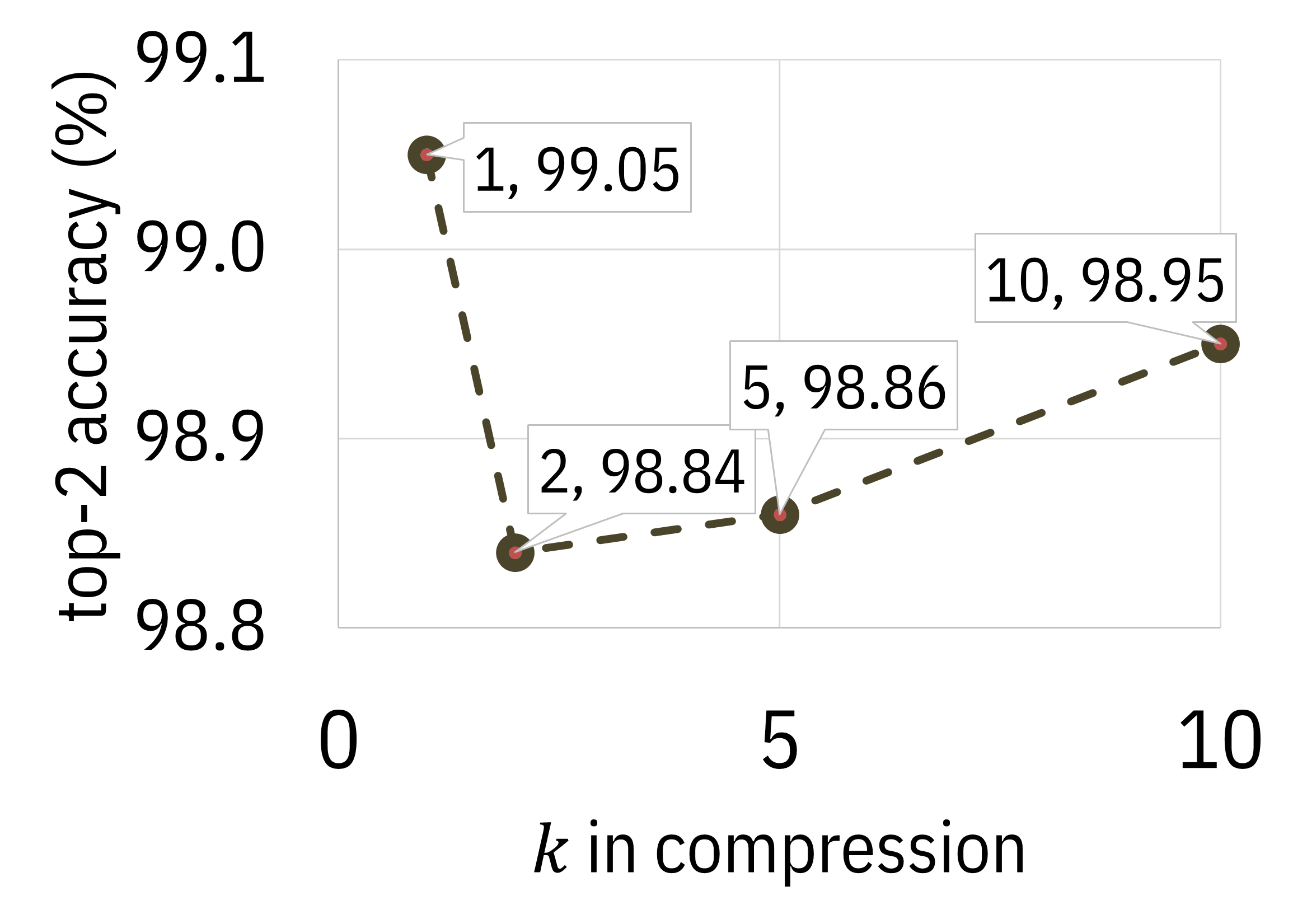}}
    ~
	\subfloat[\footnotesize CIFAR-100 aggregator Top-5\label{fig:cifar100_top5}]{%
    \includegraphics[width=0.22\textwidth]{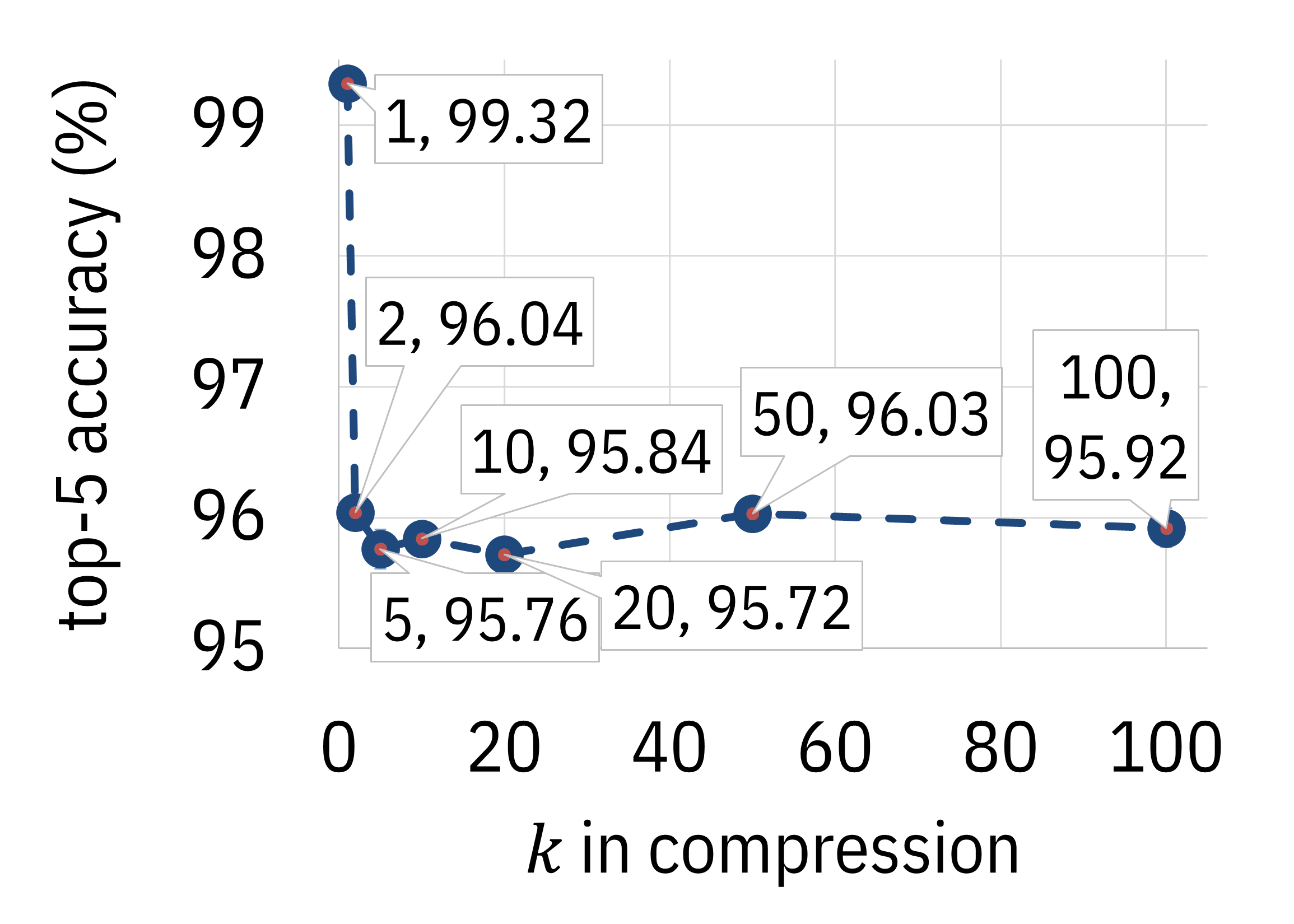}}
	~
	\subfloat[\footnotesize Food-101 aggregator Top-5\label{fig:food101_top5}]{%
    \includegraphics[width=0.22\textwidth]{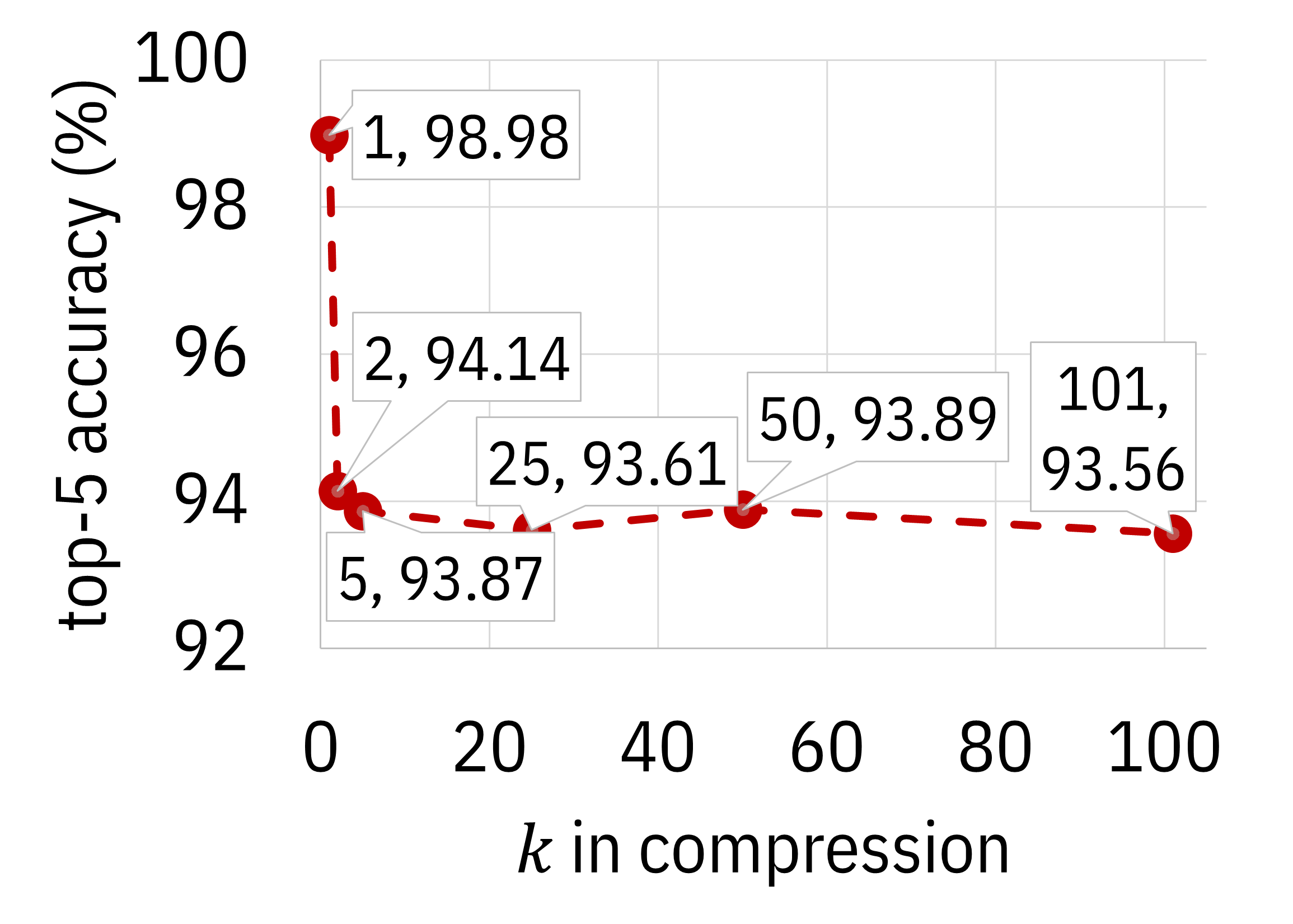}}
	~
	\subfloat[\footnotesize ImageNet aggregator Top-5\label{fig:imagenet_top5}]{%
    \includegraphics[width=0.22\textwidth]{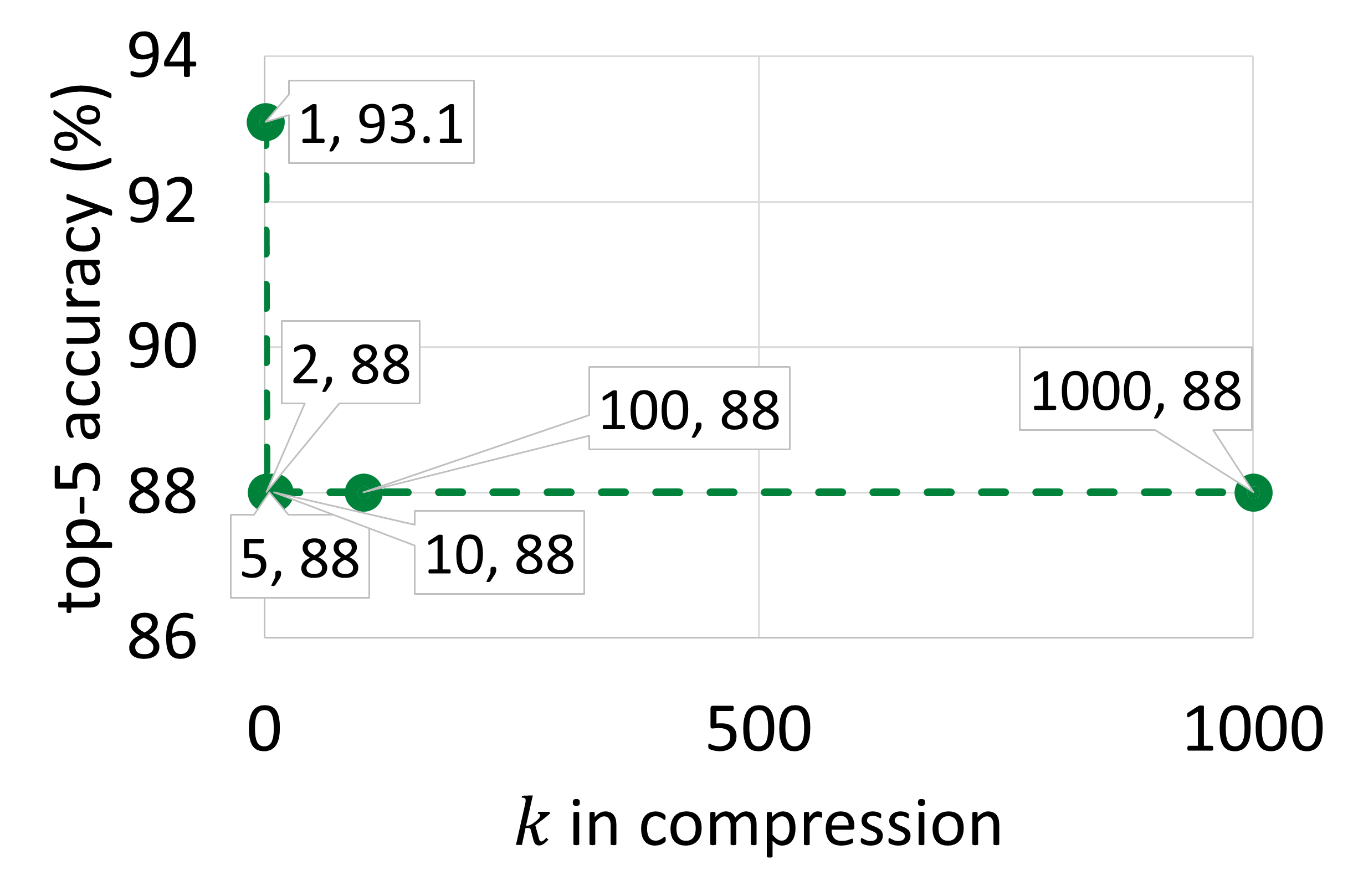}}
    % \\
    % \subfloat[\footnotesize MNIST Top-1\label{fig:mnist_top1}]{%
    %   \includegraphics[width=0.2\textwidth]{k_experiment_mnist_top1}}
    % ~
    % \subfloat[\footnotesize MNIST Top-2\label{fig:mnist_top2}]{%
    %   \includegraphics[width=0.2\textwidth]{k_experiment_mnist_top2}}
    % ~
    % \subfloat[\footnotesize F-MNIST Top-1\label{fig:fmnist_top1}]{%
    %   \includegraphics[width=0.2\textwidth]{k_experiment_fmnist_top1}}
    % ~
    % \subfloat[\footnotesize F-MNIST Top-2\label{fig:fmnist_top2}]{%
    %   \includegraphics[width=0.2\textwidth]{k_experiment_fmnist_top2}}
    % \\
    % ~
    % \subfloat[\footnotesize SVHN Top-1\label{fig:svhn_top1}]{%
    % \includegraphics[width=0.2\textwidth]{k_experiment_svhn_top1}}
    % ~
    % \subfloat[\footnotesize SVHN Top-2\label{fig:svhn_top2}]{%
    % \includegraphics[width=0.2\textwidth]{k_experiment_svhn_top2}}
    
    \caption{Impact of different $k$ values on accuracy of different benchmarks. $k$: the number of top predictions returned by each node.}
    \label{fig:k_experiment}
    \vskip -0.15in
\end{figure*}

\noindent\textbf{Ensemble Accuracy:}
% In our compression and decompression algorithm, we only select $k$ classes with highest confidence scores from each prediction vector, and send them along with their indices. 
In this experiment, we evaluate how different values of $k$ can impact the final accuracy. 
% We aggregate prediction vectors of the variants $V_1$ through $V_5$, and 
% We increase $k$ to $\max{\frac{\#classes}{2}}$.
Fig.~\ref{fig:k_experiment} illustrates a significant jump in $top\mhyphen 1$ accuracy from $k=1$ to $k=2$. For $k>2$, the accuracy does not significantly increase.
Another interesting observation is on $top\mhyphen m$ accuracy: sending the prediction with the highest confidence score per variant improves $top\mhyphen m$ accuracy. 
The reason is that when aggregating multiple prediction vectors, classes with lower scores behave like noise. By omitting these elements from our prediction vectors, we indeed achieve higher $top\mhyphen m$ accuracy. 
It appears to be a trade-off between higher $top\mhyphen 1$ and $top\mhyphen m$ accuracies. 
Apparently, $k=2$ is a good choice and we use it for the rest of our experiments.

%In Section \ref{sec:experiments,subsec:compression} we investigated the impact of different values for $k$ in compression algorithm on accuracy. Here, we study its effect on bandwidth usage. As illustrated in Table~XXTBXX,

\vspace{-1em}
\subsection{Scaling Accuracy}	\label{sec:experiments,subsec:scaling_acc}
\revise{1}{
We generated different variants each having unique characteristics. 
Here, we analyze the impact of combining different variants on the ensemble accuracy (Fig.~\ref{fig:scaling_experiment}). 
We show results on CIFAR-10, CIFAR-100, Food-101, and ImageNet. The other three datasets behave similarly to CIFAR-10. 
% Although our variants are independent and can be executed in-parallel on separate machines, for simpler comparison,
% to make comparison with the baseline easier, 
% We show the final accuracy based on the \textit{combined} MACs.
We show the final combined accuracy with respect to the average per-device MACs.
}
% TODO higlight the following line
\revise{1}{Note that the horizontal axis is the average MACs over participating variants.}
% we draw two charts per dataset: the first one compares the total number of parameters and the achievable accuracy for each combination. The second one, compares their accuracy based on the combined MACs required instead of the total number of parameters. 
\revise{1}{We also added results from the combination of the variants $V_1$ through $V_5$ ($V_7$ on Food-101 and ImageNet) while using ${k=5}$ (in the compression module) and without compression as well. To clarify, $V_i$ on the figures belonging to the results on ImageNet refers to $V_{im[i]}$.}
It is worth noting that even the combined number of parameters and MACs of our models are fewer than the baseline, except the ImageNet variants which are still less complex than the baseline.
\revise{1}{
On CIFAR-10, we almost reach the same accuracy as the baseline by combining $V_1$, $V_2$, and $V_3$. Addition of $V_4$ and $V_5$ improves the accuracy incrementally and enables us to achieve even better results than the baseline. 
On CIFAR-100, comparable accuracy can be achieved by aggregating the results of all five variants. 
}
% It is interesting that our five models altogether have $2\times$ fewer MACs than the baseline. 
The most MACs is required by $V_5$ which is ${\sim}6\times$ fewer than the original MobileNetV2.
The Food-101 dataset is rather more complex as the size of its original images are two orders of magnitude bigger than the other benchmarks. 
To achieve comparable results, we generated $V_6$ and $V_7$. 
\revise{1}{
We reach similar accuracy to the baseline by combining the first \textbf{six} variants. Inclusion of $V_7$ further improves the result, however, at the cost of more computation.
}
\revise{1}{The ImageNet dataset is far bigger than the other benchmarks both in input size and number of classes. Therefore, the variants uesd so far ($V_1$-$V_7$) do not provide reasonable accuracy. Consequently, we designed, generated and report seven new and more capable variants ($V_{im1}$-$V_{im7}$). These new variants differ in depth multiplier ($d_i{=}0.5$) and the number of neurons in the last layer, and therefore are larger and more accurate than the previous ones (see Table~\ref{table:models_characteristics}).
Similar to Food-101, we achieve an accuracy comparable to the baseline via ensemble of the first six variants. Addition of $V_{im7}$, leads to further improvement.
}

In all the benchmarks, increasing $k$ in the compression step or transmitting the predictions without any compression gives better results but the differences are not significant. 
Intrestingly, by increasing the number of variants to ${>=}6$ in the Food-101, this gap disappears. 
This can be due to the increase in the number of variants, particularly one or two more accurate ones.

\vspace{-1em}
\subsection{Effect of Every Variant on Accuracy}	\label{sec:experiments,subsec:model_counts}

% \begin{table}[b]
% 	\vskip -0.2in
% 	\caption{Characteristics of additional variants for Food-101.}
% 	\vskip -0.2in
% 	\label{table:v6_v7_characteristics}
% %	\vskip 0.05in
% 	\begin{center} \begin{small} \begin{tabular}{lccc}
% 		\hline
% 		Model	& Input Size		& \#Params	& \#MACs	\\
% 		\hline
% 		$V_6$	& $3\times256^2$	& 390k		& 71M		\\
% 		$V_7$	& $3\times320^2$	& 410K		& 112M		\\
% 		\hline
% 		\end{tabular} \end{small} \end{center}
% 	% \vskip -0.15in
% \end{table}

\begin{figure}[t]
	\centering
	\begin{minipage}{0.49\columnwidth}
	  \subfloat[\footnotesize CIFAR-10\label{fig:scaling-cifar10}]{%
		\includegraphics[width=\columnwidth]{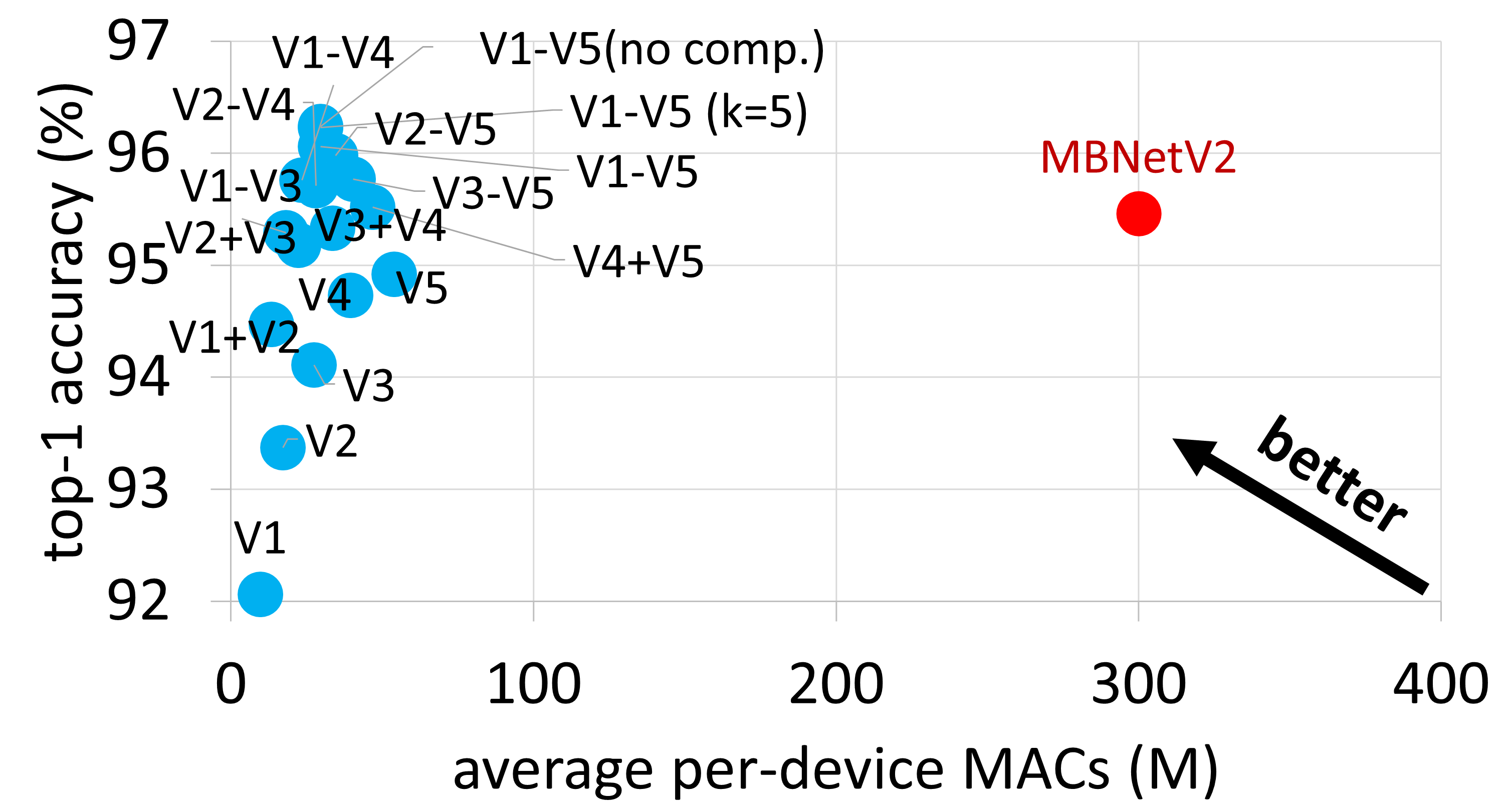}}
	  \\
	  \subfloat[\footnotesize CIFAR-100\label{fig:scaling-cifar100}]{%
	  \includegraphics[width=\columnwidth]{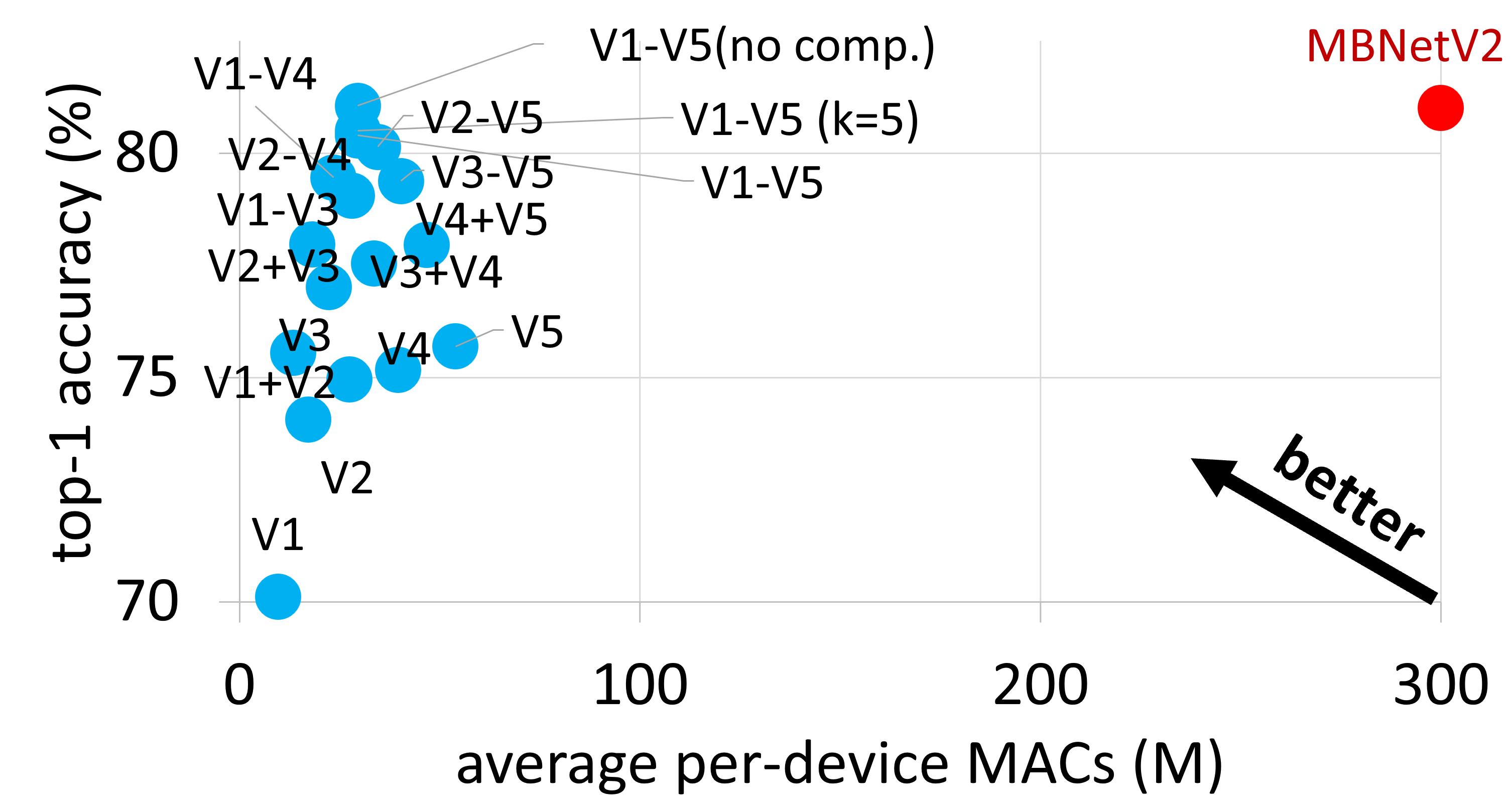}}
	  \\
	  \subfloat[\footnotesize Food-101\label{fig:scaling-food101}]{%
	  \includegraphics[width=\columnwidth]{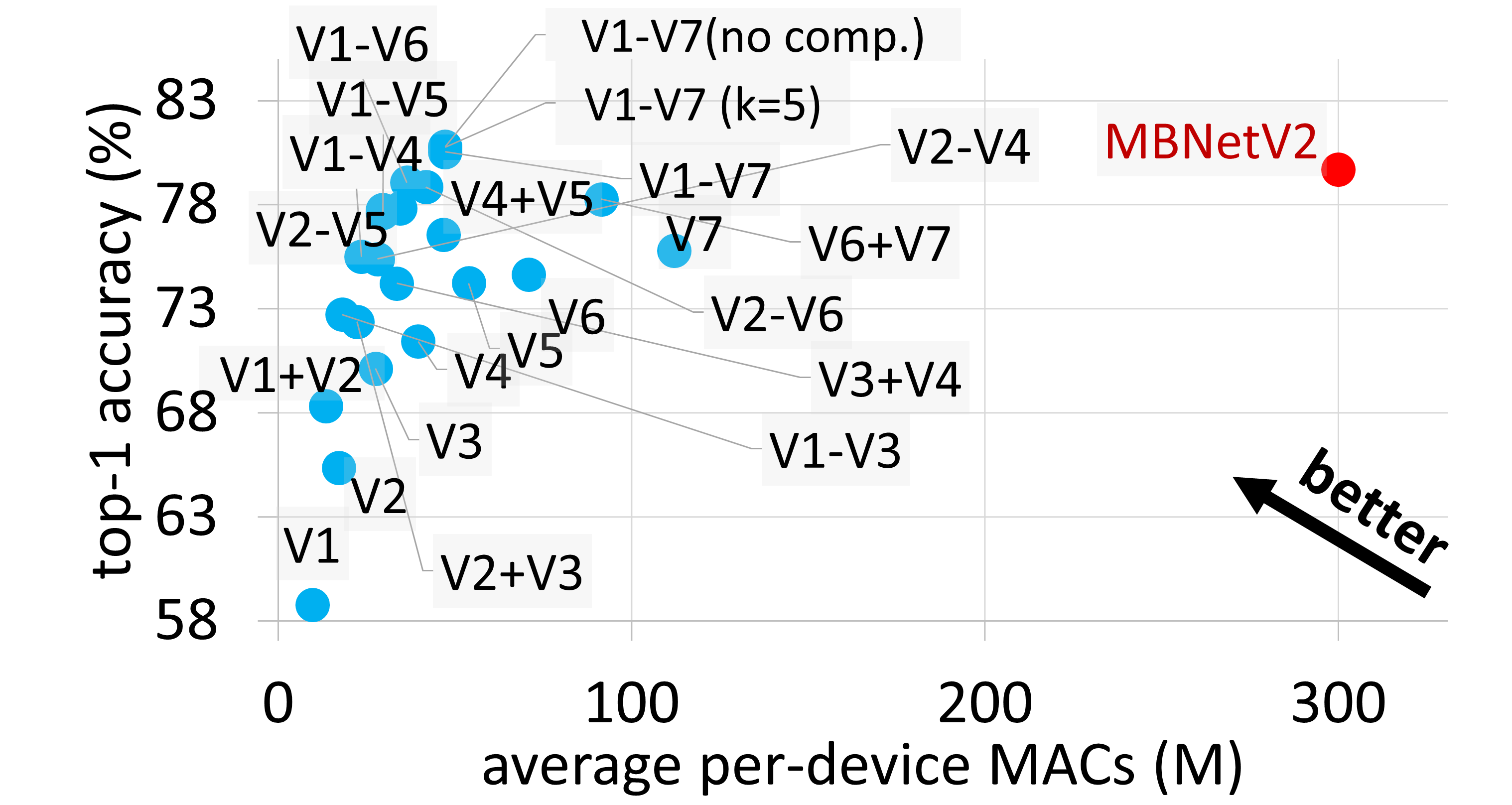}}
	  \\
	  \subfloat[\footnotesize ImageNet\label{fig:scaling-imagenet}]{%
	  \includegraphics[width=\columnwidth]{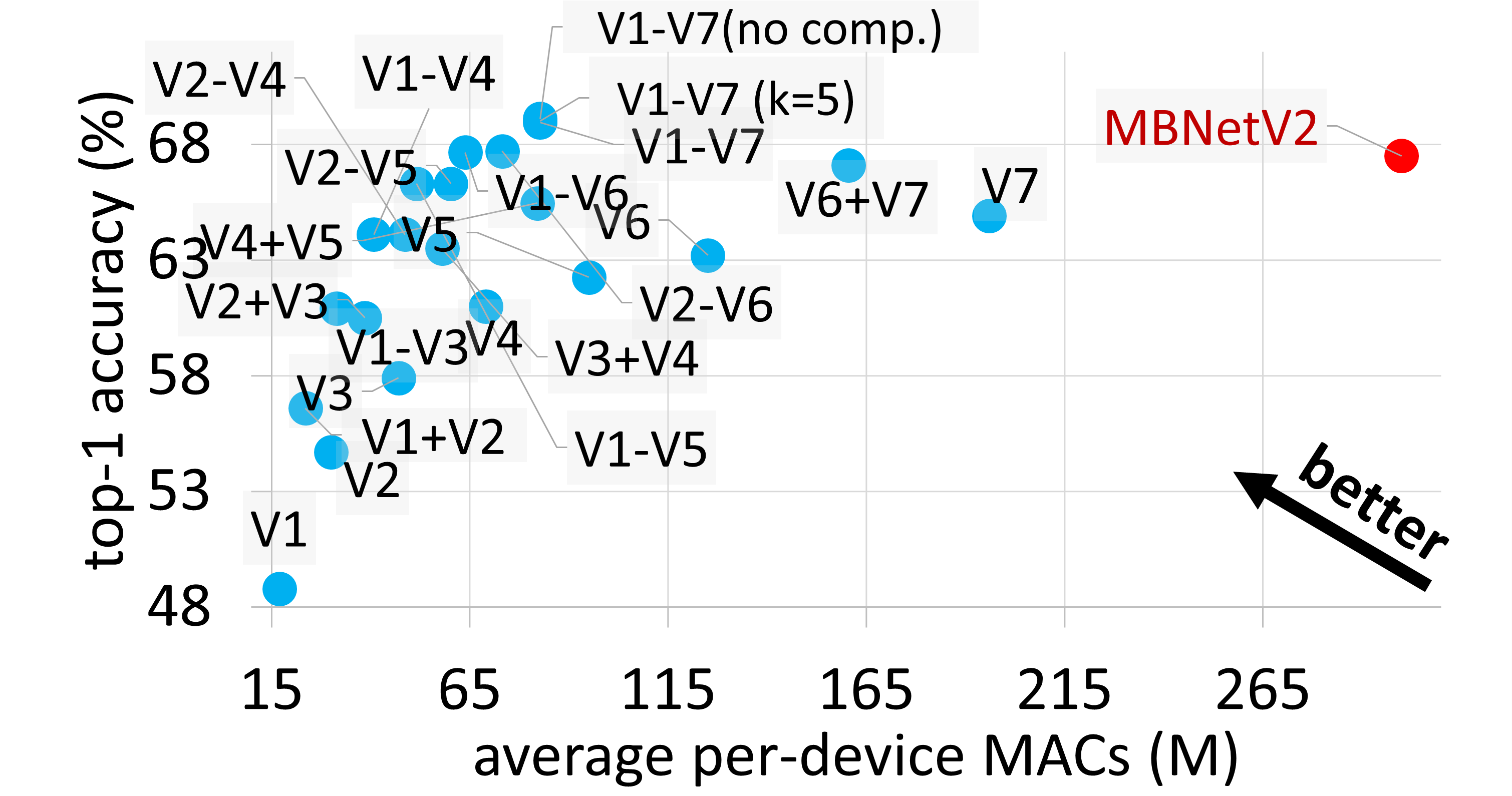}}
		\caption{Impact of each variant \revise{1}{(and different combinations)} on the final accuracy (Please zoom in).}
		\label{fig:scaling_experiment}
	\end{minipage}\hfill
	\begin{minipage}{0.49\columnwidth}
		\centering
		\subfloat[\footnotesize CIFAR-10\label{fig:scaling-x-cifar10}]{%
		  \includegraphics[width=\columnwidth]{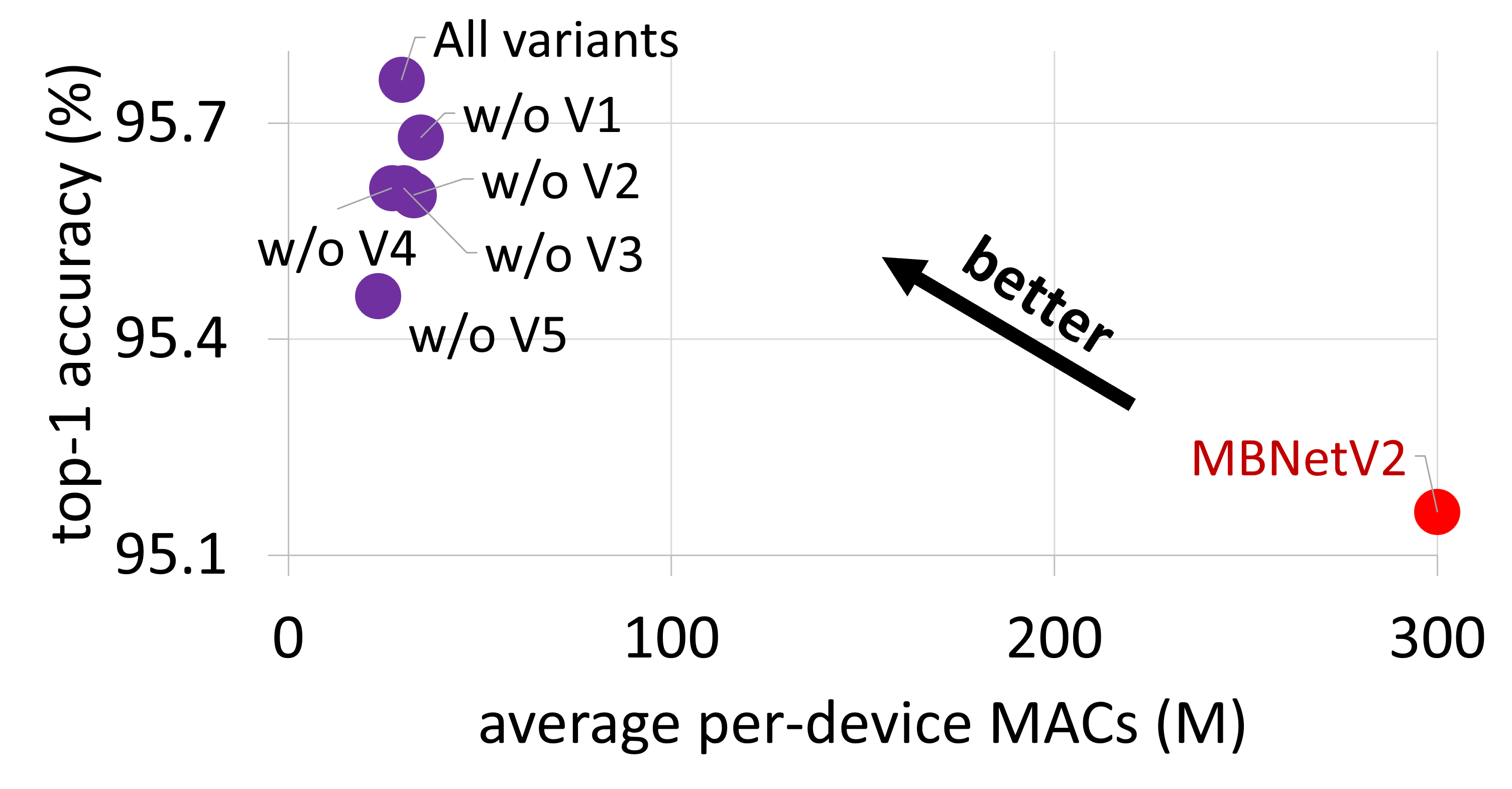}}
		\\
		\subfloat[\footnotesize CIFAR-100\label{fig:scaling-x-cifar100}]{%
		\includegraphics[width=\columnwidth]{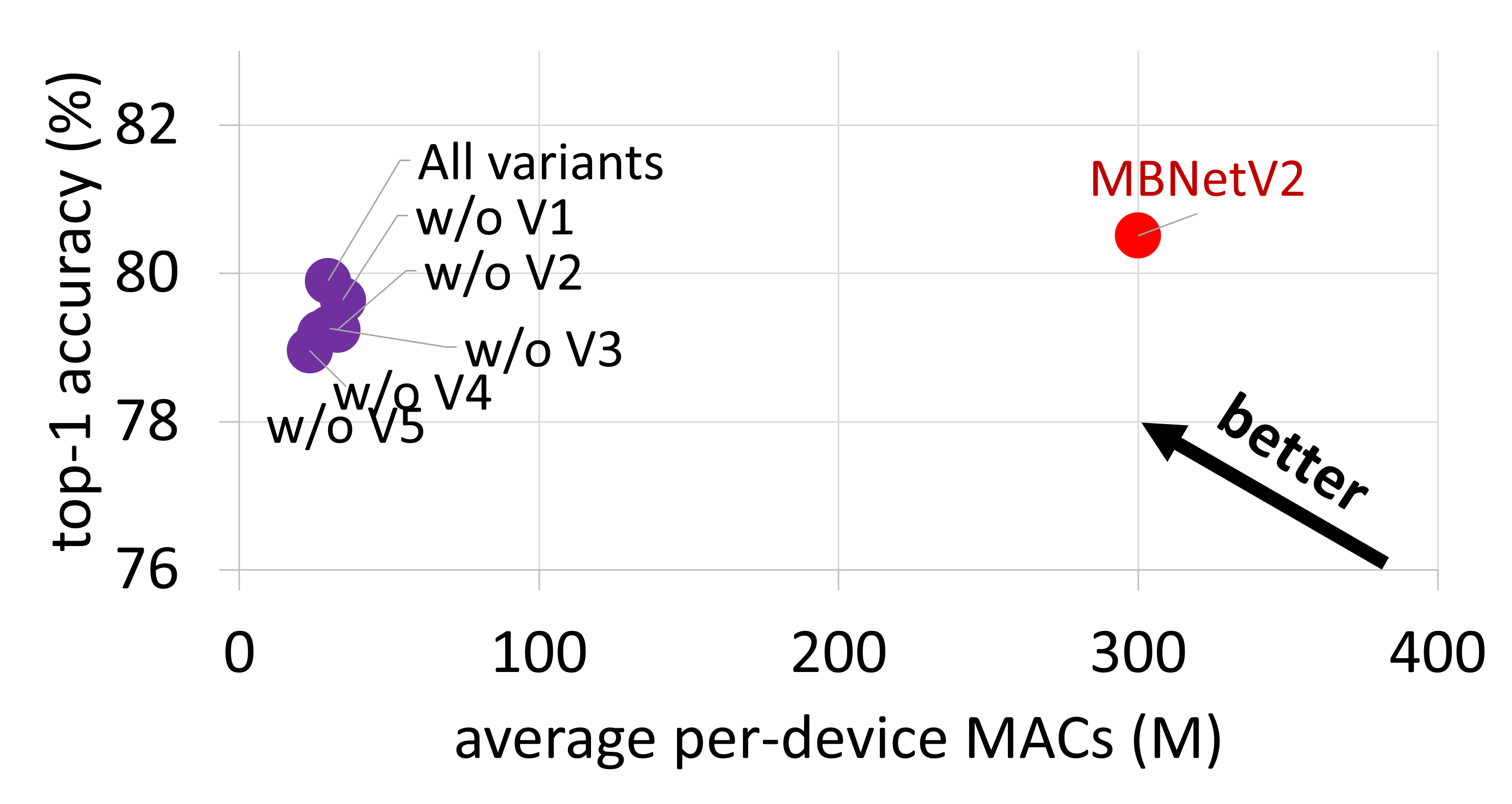}}
		\\
		\subfloat[\footnotesize Food-101\label{fig:scaling-x-food101}]{%
		\includegraphics[width=\columnwidth]{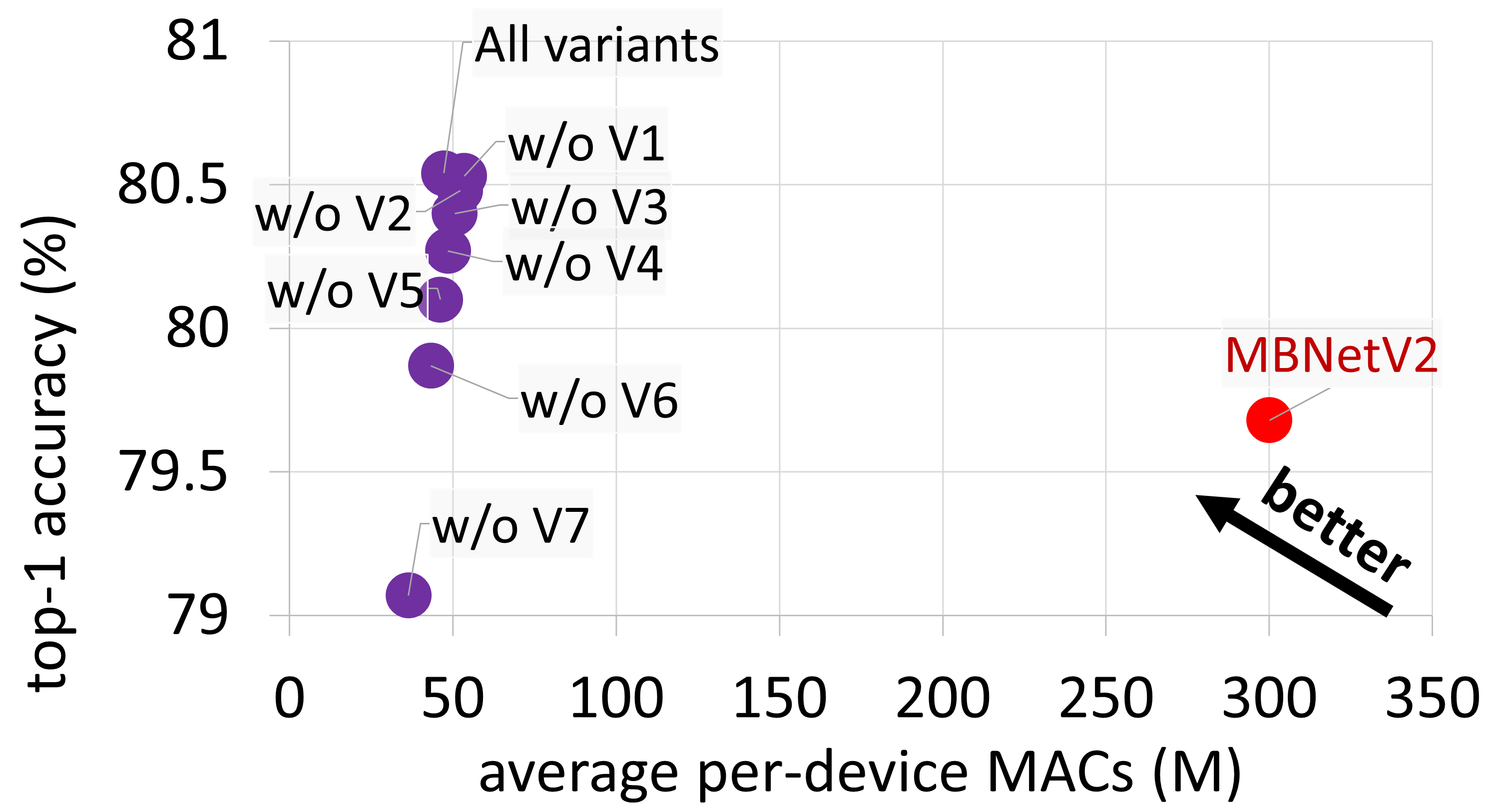}}
		\\
		\subfloat[\footnotesize ImageNet\label{fig:scaling-x-imagenet}]{%
		\includegraphics[width=\columnwidth]{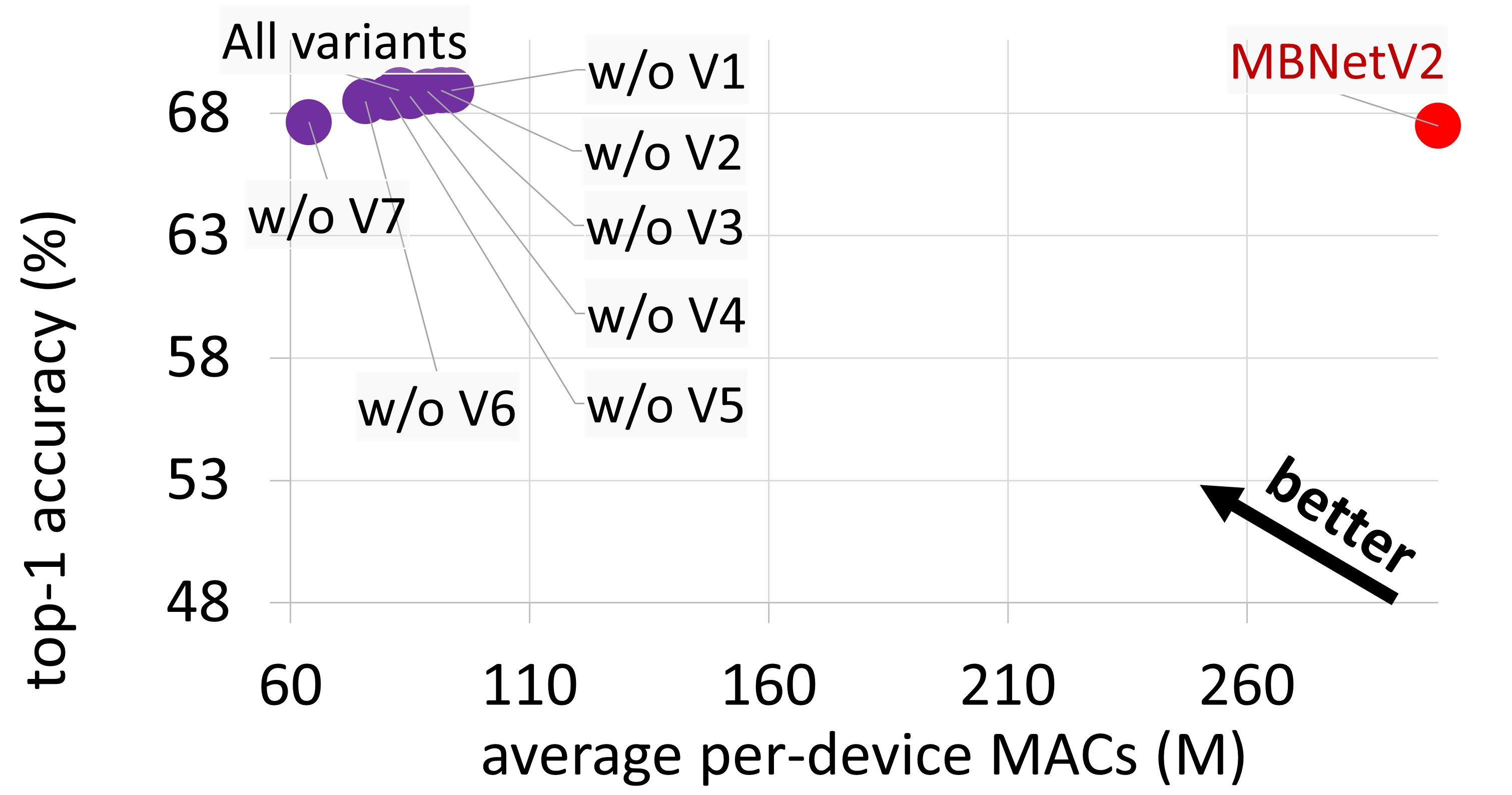}}
		\caption{How omitting a variant affects the final accuracy. (Please zoom in)}
		\label{fig:scaling_exclude_experiment}
	\end{minipage}
	\vskip -0.15in
\end{figure}

To find out whether each variant actually impacts the final accuracy, we evaluate the predictions returned by the combination of our variants while excluding only one of them. 
% Similar to Section \ref{sec:experiments,subsec:aggregate_acc}, we draw two charts for each of the candidate benchmarks. 
As shown in Fig.~\ref{fig:scaling_exclude_experiment}, in all datasets, omitting each variant has a negative impact on the final accuracy. 
This is especially interesting in less accurate variants. \revise{1}{For example, on CIFAR-100, $V_1$ has ${\sim}70\%$ accuracy itself which is more than $10\%$ lower than the baseline, yet when included in the aggregation, it can improve the final accuracy.} %by ~$0.3\%$. 
Same is true in the other benchmarks except. \revise{1}{On ImageNet, however, we do not observe this pattern. Interestingly, although addition of variants increases accuracy (Fig.\,\ref{fig:scaling-imagenet}), excluding one of them does not seem to negatively impact the final accuracy.}

% \begin{figure}[h]
%   \centering
%     \subfloat[\footnotesize CIFAR-10\label{fig:scaling-cifar10}]{%
%     %   \includegraphics[width=0.50\columnwidth]{scaling-excluding-cifar10-params}
%       \includegraphics[width=0.50\columnwidth]{scaling-excluding-cifar10-macs}}
%     \\
%     \subfloat[\footnotesize CIFAR-100\label{fig:scaling-cifar100}]{%
%     % \includegraphics[width=0.50\columnwidth]{scaling-excluding-cifar100-params}
%     \includegraphics[width=0.50\columnwidth]{scaling-excluding-cifar100-macs}}
%     \\
%     \subfloat[\footnotesize Food-101\label{fig:scaling-food101}]{%
%     % \includegraphics[width=0.50\columnwidth]{scaling-excluding-food101-params}
%     \includegraphics[width=0.50\columnwidth]{scaling-excluding-food101-macs}}
%     \caption{How omitting a variant affects the ultimate accuracy}
%     \label{fig:scaling_exclude_experiment}
%   %     \vskip -0.2in
%   \end{figure}

% TODO: Bigger fonts
	
\vspace{-1em}
\subsection{Availability in Presence of Failure}	\label{sec:experiments,subsec:availability}

One serious drawback of utilizing idle nodes is preemption in presence of higher priority jobs (e.g., their main tasks). Missing real-time deadlines due to unexpected communication or computation issues are other reasons that indicate the demand for a fault-tolerant parallelism approach. Nevertheless, \modelp{}  and \classp{}  need to get the results of all contributing machines to provide the final outcome.
% Variant parallelism in the first place aimed to be robust in case of failure of compute machines. 
By design, our architecture is fault-tolerant even if $n\mhyphen 1$ worker nodes are not able to return their predictions. 
% This is because every variant generates full-class predictions. 
% Thus, when a node does not return its vector, the master can still make a decision by combining the results received from other nodes. 
Our aggregation module also supports this behavior.
As shown in Fig.~\ref{fig:scaling_experiment}, we can still have results by contribution of a fraction of variants being executed on different machines. As the number of faulty nodes increases (i.e., fewer contributing variants), the final accuracy decreases. The achievable accuracy depends on the variants remained for the aggregation. In the worst case, all nodes fail except the one that executes $V_1$ as it has the lowest accuracy among all, which is still better than \modelp{}/\classp{}  that fail to provide any predicton. \revise{1}{Our \variantp{} can be viewed as providing a graceful degradation in presence of increasing levels of faults/preemptions in the environment.}

\vspace{-1em}
\subsection{Other Metrics: Inference Time, MACs, Model Size}	
\label{sec:experiments,subsec:other_perfs}

We report the performance of our variants on the other common metrics. We convert each variant designed for and trained on Food-101 to TFLite \cite{tflite}, and execute them for $100$ iterations. The results are illustrated in Table~\ref{table:perf_metrics}.
% Here, our aggregation module contains the decompression, scaling, and aggregation components for six variants.
Since each variant has different characteristics, the slowest one determines the eventual response time. Therefore, depending on the available combination of nodes, one can get $2.5\mhyphen 13.2\times$ speedup compared to the baseline. 
Considering a smart home scenario with a star topology and the round-trip time ${RTT=2ms}$, the estimated speedup would be $2.4\mhyphen 10.7\times$
%\subsection{Response Time}	\label{sec:experiments,subsec:latency}
%
%
%\subsubsection{Response Time Breakdown}	\label{sec:experiments,subsec:time_breakdown}

%\begin{figure}[t!]
%	\centering
%	\includegraphics[width=\columnwidth]{response_time_breakdown.png}
%	\vskip -0.1in
%	\caption{Response time break down.}
%	\label{fig:k_experiment_output_size}
%\end{figure}

\vspace{-1em}
\subsection{\revise{2}{Variant Parallelism vs. Class Parallelism}}
\revise{2}{
We concisely compare our method with the current state-of-the-art deep learning parallelism algorithm, \textit{sensAI} \cite{MLSYS2021_sensai} which is a \classp{} technique. %(Table~\ref{table:vp_cp}).
}

\noindent\textbf{Fault-Tolerance:}
% Our main objective in variant parallelism is to design a fault-tolerant system. 
After sending requests to workers, the master node waits for a time window to receive as many responses as possible. Thus, in the event of preemtion due to processing high priority tasks or an issue in a worker or the network, the master node treats them as a fault.
We demonstrated that in case of failure on even $n\mhyphen 1$ worker nodes, we can still get a response. 
In \classp{}, however, since each model predicts different class(es), all contributing machines have to successfully process and send their results to the master. 
One can improve it by executing replications on additional machines, but it increases both complexity and cost. 
Another idea is to apply overlapping class-aware pruning \cite{yang2020robust}. 
In this type of replication, in presence of failure in one machine, the results can be aggregated from other machines with accuracy drop. Both ideas increase computation and communication costs. 
% Moreover, they only mitigate the problem.

\noindent\textbf{Response Time, MACs, Model Size, Training Time:}
% The \textit{sensAI} paper reports its response time on a number of deep learning models including MobileNetV2 as an efficient architecture, which is also our baseline model. 
As depicted in Table~{\ref{table:perf_metrics}, by leveraging class parallelism, \textit{sensAI} can reach up to ${\sim}2\times$ speedup using $10$ parallel machines. 
We showed that variant parallelism gets ${\sim}2.5\mhyphen 3.2\times$ speedup compared with the baseline while requiring roughly half of in-parallel compute nodes compared to \textit{sensAI}. Further note that, we trained our variants for $135$ epochs which is $32.5\%$ less than class parallelism, yet we achieve comparable or higher accuracy. However, since our infrastructures and scheduling policies are different, direct comparison might be unfair.

\begin{table}[t]
	\vskip -0.2in
	\caption{Comparison of our variants with the baseline and \textit{sensAI}.}
	\vskip -0.2in
	\label{table:perf_metrics}
	\begin{center} \begin{small}\begin{tabular}{p{1cm}p{1.4cm}p{0.8cm}p{1.7cm}p{1.8cm}}
		\hline
		Model			& Time (ms)			 & Speedup		& \#MACs Gain	& \#Params Gain		\\
		\hline
		MBNet2		 	& $226\pm0.5$		 & $1{\times}$	& $1{\times}$	& $1{\times}$	\\
		\hline
		$V_6$		 	& $91\pm0.4$	     & $2.5\times$	& $4.3\times$	& $5.8\times$	\\
		$V_5$		 	& $71\pm0.2$	     & $3.2\times$	& $5.6\times$	& $6.1\times$	\\
		$V_4$		 	& $53\pm0.2$	     & $4.2\times$	& $7.6\times$	& $6.3\times$	\\
		$V_3$		 	& $38\pm0.2$	     & $5.9\times$	& $10.9\times$	& $6.7\times$	\\
		$V_2$		 	& $26.5\pm0.1$	     & $8.6\times$	& $17.5\times$	& $6.9\times$	\\
		$V_1$		 	& $17\pm0.05$	     & $13.2\times$	& $31\times$	& $7.1\times$	\\
		Aggregate	  	& $0.05\pm0$	     & -			& -				& -				\\
		\hline
		\textit{sensAI}	& -					 & $2\times$	& $3.5\times$	& $5\times$		\\
		\hline
		\end{tabular} \end{small} \end{center}
	% \vskip -0.2in
	\vspace{-1.5em}
\end{table}

\noindent\textbf{Output Size:}
In the tasks with few classes (e.g. CIFAR-10, SVHN, and MNIST) almost the same number of values must be transmitted. However, for the datasets with ${C>=100}$ (e.g. CIFAR-100 and Food-101), we transmit ten bytes per variant which depending on the number of clusters in \classp{}, it can be up to $2\times$ ($20\times$ on ImageNet) less than \textit{sensAI}.

\noindent\textbf{Flexibility:}
Since each variant is being executed independently, it gives us more flexibility than \classp{}  to achieve different goals. We can generate additional variants based on more than one basic architecture, or design a highly customized variant for contributing machines. We can have several variants with different characteristics to be deployed on heterogeneous compute machines. \variantp{} can also be combined with other distribution schemes including \classp{}.

\section{Conclusion \& Future Directions}
\label{sec:disscusion}

In this paper, we presented Variant Parallelism, a more flexible and fault-tolerant distribution scheme based on ensemble learning. Our evaluation on six different object recognition datasets demonstrates that our method can improve performance in number of parameters, MACs, and response time compared to the baseline and the current state-of-the-art.
We note potential direction to be considered in future works. 
We intend to extend it to more complicated tasks such as object detection and semantic segmentation. We expect more performance gain is achievable when tasks become more complicated.
Both \variantp{} and \classp{}  are based on the well-known ensemble learning techniques. Therefore, they can have similar weaknesses and strengths. For example, in \variantp{} combining variants with identical characteristics, may not help achieving significant accuracy boost unless we retrain each of them with different random seeds \cite{allen2020towards}.
Contributing machines in \classp{}  and \variantp{} observe the same data. 
In some scenarios, e.g., a smart city with long latency or bandwidth bottlenecks, transmitting input data can impact the end-to-end latency. Combining split computing paradigm 
% \cite{kang2017neurosurgeon, teerapittayanon2016branchynet, matsubara2019distilled, sbai2021cut, yao2020deep, laskaridis2020spinn} 
\cite{kang2017neurosurgeon, sbai2021cut} with \classp{}  or \variantp{} can help mitigate the problem. 
\revise{2}{On that end, a hierarchy of resources with different computation and communication characteristics (e.g., end-edge-cloud) is another direction which has recently gained attention and has its own challenges \cite{asadi2023ensemble, zhou2022joint, zhou2022reverse}.}
\revise{2}{For generating variants we apply untargeted structured pruning. Investigating targeted pruning methods (e.g., \cite{pmlr-v206-zhao23b}) to further improve retraining and inference times, therefore, can be considered as part of the future work to this paper.}

\bibliography{variant_parallelism}
\bibliographystyle{IEEEtran}

% \begin{IEEEbiographynophoto}{Navidreza Asadi}
% Navidreza Asadi
% \end{IEEEbiographynophoto}

% \begin{IEEEbiography}[{\includegraphics[width=1in,height=1.25in,clip,keepaspectratio]{fig1.png}}]{IEEE Publications Technology Team}
% In this paragraph you can place your educational, professional background and research and other interests.\end{IEEEbiography}

% \begin{IEEEbiographynophoto}{Maziar Goudarzi}
%   Maziar Goudarzi
%   \end{IEEEbiographynophoto}
  
%   \begin{IEEEbiography}[{\includegraphics[width=1in,height=1.25in,clip,keepaspectratio]{fig1.png}}]{IEEE Publications Technology Team}
%   In this paragraph you can place your educational, professional background and research and other interests.\end{IEEEbiography}

\end{document}